\documentclass[aps,prl,twocolumn,superscriptaddress,showpacs]{revtex4-2}
\usepackage{amsmath,amssymb,amsfonts,amsthm}
\usepackage{braket}
\usepackage{graphicx}
\usepackage{bm}
\usepackage{bbm}
\usepackage{color}
\usepackage[toc,page,header]{appendix}
\usepackage{minitoc}
\usepackage[subfigure]{tocloft}
\usepackage{booktabs,tabularx}
\usepackage{dcolumn}   
\usepackage{comment}
\usepackage{epstopdf}
\usepackage{bbold}
\usepackage[colorlinks=true,linkcolor=blue,urlcolor=blue,citecolor=blue]{hyperref}
\usepackage{commath}
\usepackage[caption=false]{subfig}
\usepackage{tocloft}
\usepackage{afterpage}

\begin{document}

\title{Optimal twirling depth for classical shadows in the presence of noise}
\author{Pierre-Gabriel Rozon}
\email{pierre-gabriel.rozon@mail.mcgill.ca}
\affiliation{Physics Department, McGill University, Montr\'eal, Qu\'ebec H3A 2T8, Canada}
\author{Ning Bao}
\affiliation{Northeastern University, Boston, MA USA}
\affiliation{Brookhaven National Laboratory, Upton, NY}
\author{Kartiek Agarwal}
\affiliation{Physics Department, McGill University, Montr\'eal, Qu\'ebec H3A 2T8, Canada}
\affiliation{Material Science Division, Argonne National Laboratory, Argonne, IL 60439, USA}
\date{\today}
\begin{abstract}
The classical shadows protocol is an efficient strategy for estimating properties of an unknown state $\rho$ using a small number of state copies and measurements. In its original form, it involves twirling the state with unitaries from some ensemble 
and measuring the twirled state in a fixed basis. 
It was recently shown that for computing local properties, optimal sample complexity (copies of the state required) is remarkably achieved for unitaries drawn from shallow depth circuits composed of local entangling gates, as opposed to purely local (zero depth) or global twirling (infinite depth) ensembles. 
Here we consider the sample complexity as a function of the depth of the circuit, in the presence of noise. We find that this noise has important implications for determining the optimal twirling ensemble. Under fairly general conditions, we i) show that any single-site noise can be accounted for using a depolarizing noise channel with an appropriate damping parameter $f$; ii) compute thresholds $f_{\text{th}}$ at which optimal twirling reduces to local twirling for arbitrary operators and iii) $n^{\text{th}}$ order Renyi entropies ($n \ge 2$); and iv) provide a meaningful upper bound $t_{\text{max}}$ on the optimal circuit depth for any finite noise strength $f$, which applies to all operators and entanglement entropy measurements. These thresholds strongly constrain the search for optimal strategies to implement shadow tomography and can be easily tailored to the experimental system at hand. 

\end{abstract}
\maketitle

\paragraph{\textbf{Introduction.}}
The growing number of experimental platforms \cite{ARUTE2019_Sycamore, SATZINGER2021_TopologicalSycamore, MI2021_TimeCrystalSycamore, CRUZ2019_GHZstateIBM, EBADI2022_QuEra, Willsch2022_Dwave, HAFFNER2008_IonsComputers} capable of realizing quantum states with rich structures calls for efficient ways to probe their properties. Standard techniques, such as quantum tomography~\cite{ HAAH2017_TomographyComplexity,HU2022_LocallyScrambling,BISIO2009_OptimalTomography,CRAMER2010_EfficientTomography,ARIANO2001_QuantumTomography,ARIANO2003_QuantumTomography, SMITHEY2010_EfficientTomography,LVOSKY2009_ContinousQuantumTomography, DUNN1995_FluoTomography}, encounter significant practical limitations because the sample complexity---the number of copies of the state needed to accurately predict it---grows exponentially with system size. Remarkably, as shown in Refs.~\cite{HUANG2020_ClassicalShadow, ZOLLER2019_ShadowsPrecursor, RAINER2019_CycleBenchmarking}, and related works~\cite{AARONSON2018_ShadowTomography, HUANG2022_ClassicalShadowsML, BU2022_ShadowsWithPauliInvariantEnsembles, CHEN2021_ShadowEstimation, STRUCHALIN2021_ExperimentalShadows, KUNJUMMEN2023_ShadowProcessTomography, BECKER2023_ContinuousClassicalShadows, ACHARYA2021_POVMTomography,LEVY2021_QuantumSahdowExperimental, ZHAO2021_FermionicShadows, WAN2022_matchgateFermions, SHIVAM2023_ClassicalShadows, ARIENZO2022_ShallowShadows, IPPOLITI2023_learnability, WHITE2023_ShadowBath, ZHANG2021_Experimental, GARCIA2021_ShadowScrambling, ELBEN2023_randomizedToolbox}, one can employ a novel protocol---the classical shadows protocol---to estimate $M$ physical properties of the system 
with a system-size independent sample complexity for operators of finite weight, and that scales only logarithmically with $M$. 


In its original form, the protocol involves rotating the state $\rho$ of the system using unitaries $U$ randomly sampled from a set $\mathcal{U}$, and subsequently measuring in a fixed basis. The resultant state $\ket{b}$ is used to create a `classical snapshot' of the state of the system, $U^\dagger \ket{b} \bra{b} U$. An appropriately weighted sum of these snapshots then serves as a corrupted estimate $\mathcal{M} (\rho)$ of $\rho$, and may be inverted (provided $\mathcal{M}$ is tomographically complete) to yield an estimate $\hat{\rho}$ of the initial state  from which various properties can be computed. The sample complexity needed to estimate these properties is determined by the so-called shadow norm (to be defined below).  

Understanding how the set $\mathcal{U}$ affects the sample complexity is of crucial importance for the practical implementation of the protocol---for certain global observables, it is useful to globally twirl the system, while for local observables, local twirling is efficient~\cite{HUANG2020_ClassicalShadow}. Remarkably in the latter case, it was observed numerically~\cite{AKHTAR2023_TNShadows, BERTONI2022_LowDepthShadows}, and explained analytically~\cite{IPPOLITI2023_OperatorSpreading}, that optimal sample complexity is obtained when $U$ are chosen from shallow depth random circuits. Specifically, to estimate an operator of weight $k$ (operating non-trivially on $k$ sites) in a system of qubits, the sample complexity can be improved from $3^k$ to $\sim 2^k$ by going from circuits of depth $0$ to depth $t_* \sim \ln(k)$, but deteriorates again upon increasing circuit depth. This non-monotonic behavior can be explained~\cite{IPPOLITI2023_OperatorSpreading} by noting a relation that connects operator weight to the shadow norm, and the competition between bulk relaxation of this weight due to random unitary dynamics and increase due to operator spreading. 


Given the immense reduction in sample complexity obtained by adopting shallow depth circuits, it is pertinent to ask how these results are affected by noise. In this letter, we study these questions by modeling the presence of single-qubit noise in the twirling circuit, and determining various bounds on the optimal depth of the circuit, and thus, the optimal ensemble $\mathcal{U}$ for computing observables 
and $n^{\text{th}}$ order Renyi entropies. 

\begin{figure}
  \centering \includegraphics[width=0.35\textwidth]{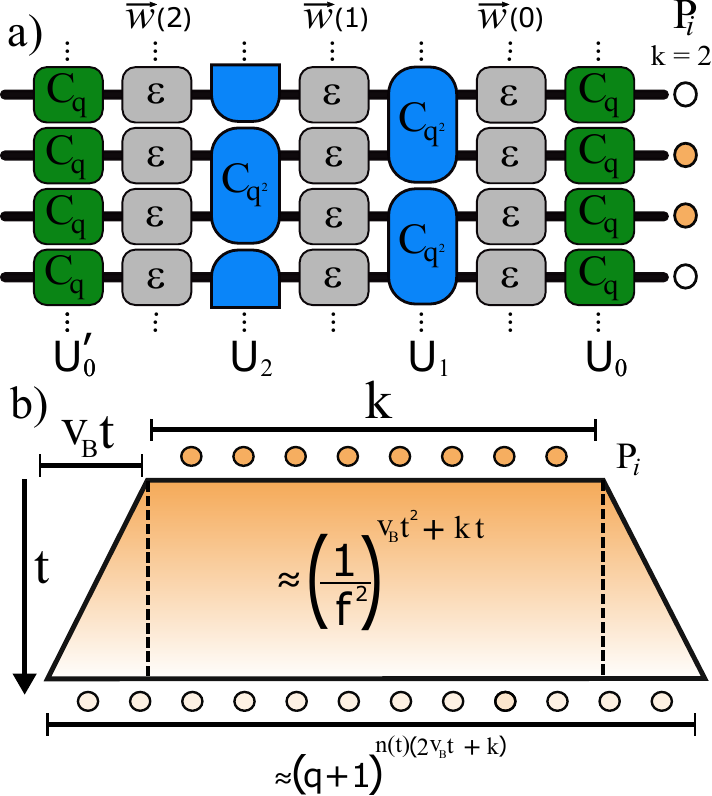}
  \caption{(a) Twirling circuit---two-qudit entangling unitaries of depth $t$ are arranged in a brickwork fashion, punctuated by single qubit noise $\varepsilon$, and sandwiched by single-qudit twirling layers. 
  (b) Mean field dynamics---a contiguous Pauli operator $P_i$ with initial weight $k$ spreads in time with a butterfly velocity $v_B$, and undergoes relaxation in the bulk to an average weight $1- 1/q^2$ due to random unitary dynamics. Depolarizing noise maintains operator weight but contributes a factor $1/f^2$ to the shadow norm at every point in spacetime with nontrivial weight. The shadow norm after depth $t$ is approximately a product of factors $\left(\frac{1}{f^2}\right)^{v_Bt^2 + kt}$ due to noise, and $(q+1)^{n(t)(k+2v_Bt)}$ due to the Pauli weight at the last layer.}
  \label{fig:Circuit}
\end{figure}

Specifically, we model the twirling unitaries $U$ using brickwork circuits of entangling two-qubit gates of finite depth $t$, punctuated uniformly by single qubit noise $\epsilon$, and sandwiched between two layers of local single-qubit twirling gates; see Fig.~\ref{fig:Circuit}. The unitary gates may be drawn from any $n \ge 2$-design unitary ensemble, e.g. the Clifford group. Importantly, as we show, the shadow norm, and thus sample complexity, depends only on a single parameter $0 < f < 1$ that characterizes the averaged rate of depolarization of Pauli operators (and their qudit generalizations) and which can be readily computed from more accurate descriptions of the noise afflicting the qudits. This permits direct interpretation of our findings by experimentalists working with specific qudit systems. 

One may expect that increasing noise makes increasing circuit depth to obtain lower sample complexity less viable. Indeed, in addition to bulk relaxation of operator weight due to random unitary dynamics, which leads to a reduction in shadow norm~\cite{IPPOLITI2023_OperatorSpreading}, a bulk competing effect characterized by the noise parameter $f$ emerges that increases the shadow norm. For sufficiently large noise, or small $f$, the improvement in sample complexity due to even one layer of entangling gates is overcome by the concomitant deterioration due to noise. A key result of this work is the computation of these thresholds for arbitrary observables and $n^{\text{th}}$ order Renyi entropies, and qudit dimension $q$. For any $f < f_{\text{th}} (q,n)$, optimal sample complexity is obtained with local twirling, regardless of the weight of the observable being estimated.



For noise below these thresholds, or $f > f_{\text{th}}$, it is useful to twirl by finite depth circuits to reduce sample complexity. To this end, we provide an upper bound $t_{\text{max}}$ on the optimal circuit depth for measurements of all observables and Renyi entropies, which we find only depends on the noise parameter $f$. The existence of a finite $t_{\text{max}}$, in contrast with the unbounded $t_* \approx \ln{k}$ in the noiseless limit, is due to the competition between noise and operator relaxation, both bulk effects, while operator growth, an edge effect, can only overcome the bulk operator decay on time scales scaling with $k$. We estimate this $t_{\text{max}}$ using matrix product state (MPS) simulations of vectors encapsulating the time evolution of the averaged operator weight in space. We find that even for small noise $f \approx 1$, the optimal circuit depth $t_*$ is upper bounded by a small value---for instance $t_{\text{max}} = 3$ for qubits with $f = 0.99$. Even for these small circuit depths the optimal sample complexity scales significantly slowly with operator/subsystem size $k$---nearly $2^k$ for dense Pauli operators and second Renyi entropy measurements as opposed to $3^k$ with just local twirling---and is thus practically advantageous.

We supplement these results with a simple mean field analysis of operator dynamics following Ref.~\cite{IPPOLITI2023_OperatorSpreading}, which suggests the optimal depth $t_*$ scales as $-\ln \left( B + C/k\right)$, where $B, C$ are system specific constants, with $B \rightarrow 0 $ for $f = 1$, and reproduces the finite $k$-independent optimal depth bound $t_{\text{max}}$ in the large $k$ limit for $f < 1$. Although this simplified analysis captures the general scaling behavior, it is not accurate enough to predict sample complexity at any finite noise and circuit depth; in forthcoming work, we discuss a more involved mean field calculation that allows for a more accurate computation of the sample complexity.

\paragraph{\textbf{Shadow norm in the presence of noise.}}We begin by discussing the shadow norm, which characterizes the sample complexity needed to accurately measure properties of the system.
Following Ref.~\cite{IPPOLITI2023_OperatorSpreading}, we study twirling unitaries  $\mathcal{U}(t)$ drawn from an ensemble invariant under local scrambling~\cite{HU2022_LocallyScrambling}---the measure $dU(t)$ remains invariant under local twirling $U(t) \rightarrow U(t)V$, $U(t) \rightarrow VU(t)$ for any tensor product of single-site Clifford unitaries $V = \bigotimes_{j=1}^Lv_j$, $v_j \in \text{Cliff}(q)$. This is explicitly accounted for by a layer of single site scrambling unitaries at the beginning and end of the circuit. Between these layers, we assume two qubit gates in a brickwork fashion drawn from a unitary $n$-design on two qubits, with $n \ge 2$. Noise is captured by introducing single qubit error channels applied between the unitary layers;  see Fig. (\ref{fig:Circuit}). The shadow channel $\mathcal{M}(\rho)$ is given by


\begin{align}\label{eq:FullChannel}
\sum_b \int_{\mathcal{U}(t)}  \underbrace{dU(t) \bra{b}K_{\varepsilon, U(t)}(\rho)\ket{b}}_{\text{Probability}(b|U(t),\rho,\varepsilon)} \underbrace{\left(U(t)^{\dagger}\ket{b}\bra{b}U(t)\right)}_{\sigma_{U(t),b}}
\end{align}

and constitutes an average over all possible measurement outcomes $\ket{b}$ and random twirling unitaries $U(t) = U_0'U_t...U_1U_0$. Here $K_{\varepsilon, U(t)}(.)$ denotes the noisy twirling channel through which the initial state $\rho$ passes before measuring in a fixed basis of states spanned by $\ket{b}$. The above choice of the snapshot $\sigma_{U(t),b}$ is standard~\cite{KOH2022_ClassicalShadowsWithNoise, JNANE2023_NoisyShadows, SHAO2023_groupShadowNoise, CHEN2021_ShadowEstimation, WU2023_NoisyFermionicShadows} but not unique. For at least uniform depolarizing noise, we find that other reasonable choices of $\sigma_{U,b}$ that assume knowledge of noise properties, do not in fact improve the shadow norm (see SM); a more rigorous analysis is merited and left for future work.  
The invariance under local twirling implies (see SM) that all Pauli operators are eigenstates of the shadow channel $\mathcal{M}$, with $\mathcal{M}(P_{i}) = \beta_{i,\varepsilon}P_{i}$. For qudits with local Hilbert space dimension $q$, we make use of generalized Pauli matrices~\cite{GHEORGHIU2014_GeneralizedPaulis} $X_j^{n_j}$, $Z_j^{m_j}$ with integers $0 \le n_j,m_j \le q-1$. 
In the absence of noise, the shadow norm \cite{IPPOLITI2023_OperatorSpreading, BERTONI2022_LowDepthShadows, AKHTAR2023_TNShadows, BU2022_ShadowsWithPauliInvariantEnsembles} $\norm{P_{i}}_{\text{sh}}^2$ is given by $\beta_{i}^{-1}$ and is related to the probability distribution of the weight vector $\vec{w}$ of the Pauli operator after twirling (the $i^{\text{th}}$ component of $\vec{w}$ is $0$ if the Pauli operator is identity at site $i$, and $1$ otherwise). In the noiseless case, if one starts with an initial Pauli operator with weight $\vec{w}(0)$, then after each unitary layer the Pauli operator has a different weight, creating a sequence $\vec{w}(0), \vec{w}(1),..,\vec{w}(t)$ which occurs with probability $\text{Pr}(\vec{w}(0),...,\vec{w}(t))$. 
Defining the marginal probability distribution $\text{Pr}(\vec{w}(t)) = \sum_{\vec{w}(1),...,\vec{w}(t-1)}\text{Pr}(\vec{w}(0),...,\vec{w}(t))$, it can be shown~\cite{IPPOLITI2023_OperatorSpreading} that
\begin{equation}\label{eq:NoiselessEigenvalue}
\beta_i = \sum_{\vec{w}(t)}\text{Pr}(\vec{w}(t))\frac{1}{(q+1)^{\norm{\vec{w}(t)}^2}}.
\end{equation}
Note that Eq. (\ref{eq:NoiselessEigenvalue}) only depends on the final weight configuration $\vec{w}(t)$. With noise applied between each unitary layer, the eigenvalue becomes sensitive to the entire history of weight vectors $\vec{w}(0),...,\vec{w}(t)$. Assuming depolarizing noise characterized by strength $f$ (which we show later captures more general single-site noise channels), each time a Pauli operator of weight $\vec{w}$ is encountered, a factor of $f^{\norm{\vec{w}}^2}$ is acquired in $\beta_{i, \epsilon}$, the eigenvalue of the shadow channel in the presence of noise. Thus the noisy eigenvalues $\beta_{i,\varepsilon}$ is given by 
\begin{align}\label{eq:numericalEq}
\left(\sum_{\vec{w}(1),...,\vec{w}(t)}\text{Pr}(\vec{w}(0),...,\vec{w}(t))f^{w_{\text{tot}}}\right)\frac{1}{(q+1)^{\norm{\vec{w}(t)}^2}}
\end{align}
where $w_\text{tot} = \sum_{i=0}^t\norm{\vec{w}(i)}^2$ is the total weight of the trajectory. The shadow norm in the presence of noise is in fact determined by both shadow channel eigenvalues $\beta_i, \beta_{i,\epsilon}$ of the Pauli operator computed in the absence/presence of noise. In particular (see SM),  
\begin{equation}
\norm{P_i}^2_{\text{sh}} = \frac{\beta_i}{(\beta_{\varepsilon,i})^2}.
\end{equation}
Finally, non-linear functions of $\rho$ can be computed via operators $O$ acting on the tensor product  $\bigotimes_{s=1}^n\rho_s$ of $n$ copies of the state, with $\rho_s = \rho$. For $O = \sum_{i_1,...,i_n}s_{i_1,...,i_n} \otimes_{j=1}^n P_{i_j}$ where $s_{i_1,...,i_n}$ are coefficients and $P_{i_j}$ are Pauli operators with support on the $j^{\text{th}}$ copy, the shadow norm (see SM) is an appropriately weighted sum of the product of shadow norms of the individual $P_{i_j}$ less the mean. That is, the shadow norm $\norm{O}^2_{\text{sh}}$ is given by \footnote{This particular form of the shadow norm assumes a specific type of estimator, but other choices can be made, and are discussed in the SM.}
\begin{align}\label{Eq:ShadowNormArbitraryO} \sum_{i_1,...,i_n}|s_{i_1,...,i_n}|^2\prod_{s=1}^n\frac{\beta_{i_s}}{\left(\beta_{\varepsilon, i_s}\right)^2}  
-\operatorname{Tr(\bigotimes_{s=1}^n\rho_s O)}^2.
\end{align}

\paragraph{\textbf{Reduction to depolarizing noise.}} We limit ourselves to the simplest kind of noise describable as a tensor product of identical single-site quantum error channels $\varepsilon = \bigotimes_{j=1}^L \varepsilon_j$  
that are time-independent, and gate-independent Markovian processes. The action of $\varepsilon_j$ can be surmised from its effect on Pauli operators---
\begin{align}\label{eq:NoiseModel}
\varepsilon_j \left(X_j^nZ_j^m \right) = \sum_{n',m'}f_{(n,m),(n',m')}X_j^{n'}Z_j^{m'}
\end{align}
where $f_{(n,m),(n',m')}$ are constrained by the requirement that $\varepsilon_j$ is a trace-preserving, completely positive map.

We find (see SM) that all off-diagonal elements of the tensor $f_{(n,m),(n',m')}$ can be disregarded in the computation of the shadow norm, while the diagonal components $f_{(n,m),(n,m)}$ only enter via an effective depolarizing strength   
$f \equiv \sum_{(n,m) \neq (0,0)}\frac{f_{(n,m),(n,m)}}{q^2 - 1}$. This is a direct consequence of the expression for the shadow channel eigenvalue 
\begin{align}
\beta_{i, \epsilon} = \int_{\mathcal{U}(t)} dU(t)\underbrace{\bra{0}K_{\varepsilon, U(t)}(P_i) \ket{0}}_{\text{Noisy}}\underbrace{\bra{0} U(t)P_i^{\dagger} U(t)^\dagger \ket{0}}_{\text{Noiseless}} \nonumber
\end{align}
which involves an average over the unitary ensemble $\mathcal{U} (t)$ of the product of both the noisy and the noiseless twirling channel applied to $P_i$. Instances where the error channel $\varepsilon_j$ maps the Pauli operator locally to a different operator thus average to zero. Moreover, the twirling of the diagonal terms leads to an averaging over the action of depolarizing noise over all non-identity $q^2-1$ local Pauli operators. The average is equally weighted because all single site Pauli operators are equally likely upon twirling. Thus, only diagonal elements of the channel $\varepsilon$ contribute, and can be captured by a single effective parameter $f$. 


\paragraph{\textbf{Noise thresholds $f_{\text{th}}$. }}The presence of noise hinders an observer's ability to extract information about the unknown state $\rho$, which is manifested here by the emergence of noise thresholds, $f_{\text{th}}$---for stronger noise, any application of the noisy quantum circuit only increases the sample complexity instead of reducing it. For a given type of measurement, these thresholds are determined by comparing the sample complexity at $t=0$ and $t=1$, which become equal at $f = f_{\text{th}}$. 

We start by computing thresholds tailored to the $n^{\text{th}}$ order Renyi entropy of a contiguous sub-region $A$ given by
$-\frac{1}{1-n}\log(\operatorname{Tr}(S_{A,n}\bigotimes_{s=1}^n\rho_s))$
where $S_{A,n}$ is the subsystem SWAP operator which acts as $S_{A,n}\ket{i_1,i_2,...,i_n}_A\otimes\ket{j_1,...,j_n}_{A^c} = \ket{i_n,i_1,...,i_{n-1}}_A\otimes\ket{j_1,...,j_n}_{A^c}$ on computational basis states, with $A^c$ the complement of $A$.
We thus see that the shadow norm for Renyi entropies may be extracted by computing the shadow norm of the SWAP operator $S_{A,n}$. By substituting the exact coefficients for $S_{A,n}$ in Eq. (\ref{Eq:ShadowNormArbitraryO}), we find closed-form expressions for the shadow norm at $t= 0, 1$, which yields the thresholds $f_{\text{th}, n}$; see Tab.~\ref{tab:fvalues}. Note that $f_{\text{th}, n}$ is not monotonic in $q$ or $n$. This is due to the structure of the Pauli decomposition of the SWAP operator (see SM), whose effects are attenuated at large $n$ or large $q$. Focusing on single Pauli operators with $k$ contiguous occupied sites, we find (see SM) that the $f_{\text{th}}$ satisfies $f_{\text{th}}^6 (f_{\text{th}} (q-1)+2)^2 - (q^2+1) = 0.$
\begin{table}
\begin{tabular}{|c|c|c|c|c|c|}
\hline
 \text{} &\text{Arbitrary operators} & \text{n = 2} & \text{n = 3} & \text{n = 4} & \text{n = 5} \\
 \hline
 \text{ q = 2 }& 0.9153 & 0.9396 & 0.9564 & 0.9535 & 0.9539 \\
 \text{ q = 3 }& 0.9349 & 0.9462 & 0.9551 & 0.9545 & 0.9545 \\
 \text{ q = 4 }  & 0.9478& 0.9544 & 0.9598 & 0.9596 & 0.9596 \\
 \text{ q = 5 }& 0.9565 & 0.9609 & 0.9646 & 0.9645 & 0.9645 \\
 \text{ q = 6 }& 0.9629 & 0.9659 & 0.9686 & 0.9686 & 0.9686 \\
 \hline
\end{tabular} 
\caption{Values of $f_{\text{th}}$ for the $n^{\text{th}}$ order Renyi entropy as a function of the qudit dimension $q$. Also displayed is the lower bound on $f_{\text{th}}$ for any operator as a function of $q$.}
\label{tab:fvalues}
\end{table}
The solutions of this equation are displayed in Tab.~\ref{tab:fvalues}, which shows that $f_{\text{th}}$ increases with $q$. This is expected, as with a larger local qudit dimension $q$, there are more possibilities for non-identity Pauli operators on any given site, which mitigates the ability of the unitary matrices to reduce the operator weight. We note that the thresholds $f_{\text{th}}$ obtained for contiguous Pauli operators are independent of $k$. The obtained thresholds are in fact lower bounds for any sparse Pauli operator since the shadow norm grows faster due to operator spreading at multiple boundaries, and therefore all observables by extension due to the completeness of the Pauli operator basis. 




\begin{figure}
  \centering \includegraphics[width=0.49\textwidth]{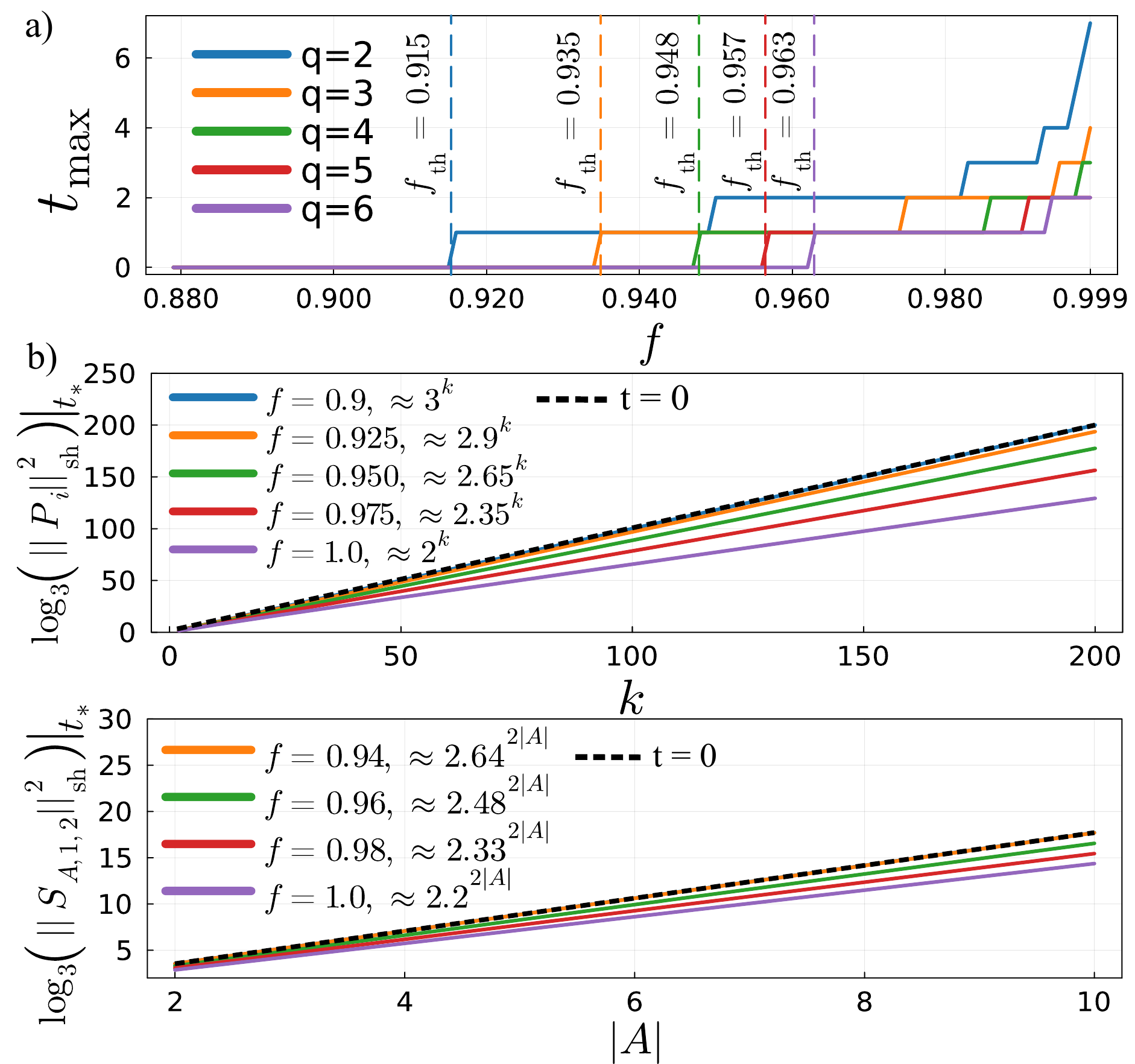}
  \caption{a) Upper bound $t_{\text{max}}$ on the optimal twirling depth $t_*$ as a function of $f$, for various choice of the qudit dimension $q$. Also displayed are the lower bound on the noise thresholds $f_{\text{th}}$. b) Scaling of the shadow norm as a function of $k$ for Pauli operators (top panel) and as a function of the sub-region size $|A|$ for the second Renyi entropy (bottom panel) when evaluated at the minimum $t_*$. }
  \label{fig:scaling}
\end{figure}


\paragraph{\textbf{Bound $t_{\text{max}}$ on optimal depth.}}When the effective noise parameter $f$ lies in the range $f \in [f_{\text{th}}, 1]$, it is useful to understand how the optimal circuit depth $t_*$ scales as a function of operator size $k$ and noise parameter $f$. Here we first estimate an upper bound $t_{\text{max}}$ on $t_*$ in the presence of noise. There are three processes relevant to computing the shadow norm---i) increasing operator weight due to opreator spreading, ii) decreasing operator weight due to bulk relaxation due to random unitary dynamics, from weight $1$ to the average weight $1-1/q^2$, and iii) factors of $1/f^2$ at any point in spacetime with a non-trivial Pauli operator, originating from noise. To obtain an upper bound on $t_*$, we can ignore operator spreading (as it would decrease $t^*$); the remaining two processes are bulk effects, and thus the $t_{\text{max}}$ obtained is independent of initial operator weight $k$. (In the noiseless case, this assumption would lead to a trivial upper bound with $t_{\text{max}} = \infty$, consistent with $t_* \sim \ln k$ scaling~\cite{IPPOLITI2023_OperatorSpreading}.)

To estimate $t_{\text{max}}$ in the presence of noise,  
we time-evolve a Matrix Product State (MPS) encoding the probability of obtaining a given weight vector $\vec{w}(t)$ times the appropriate noise attenuation factor $f^{w_{\text{tot}}}$, as in Eq.~(\ref{eq:NoisyEigenvalueRaw}), see SM. We evolve a contiguous Pauli operator of size $L$, in a system of size $L \gg t^*$, with periodic boundary conditions; the value of $t_*$ found in this context corresponds to $t_{\text{max}}$ as there is no operator growth in this setting.  We provide in Fig. \ref{fig:scaling} the value of $t_{\text{max}}$ as a function of $f$ for various $q$. As $q$ is increased, $t_\text{max}$ is further reduced, as a larger value of $q$ suppresses the effect of bulk decay. Interestingly, the upper bound on the optimal depth grows slowly with $f$ (for $f = 0.99$, we find $t_{\text{max}} = 3$), indicating that for any realistic implementation of the shadow protocol, small depth circuits reach optimality. 

\paragraph{\textbf{Mean-field dynamics; $t_*$ and optimized shadow norm scaling with $k$.}}By studying the interplay of the three above mechanisms via a simple mean-field analysis (see SM), we find a scaling $t_* \approx \frac{1}{\gamma}\ln \frac{1}{B + C/k} + o(\ln\ln \frac{1}{B + C/k})$ for $k \gg t_*$,  $t_* \gg 1$ with noise dependant constants $B,C$, where $B$ arises from the local competition of the noise $f$ with the bulk decay rate $\gamma$, while $C$ corresponds to the competition of the operator growth $v_B$ with the bulk decay. In the limit $k \rightarrow \infty$, this recovers a constant $k-$independent optimal depth. 

The mean field approach gives approximately correct scaling of the optimal depth, but is not accurate enough to predict the optimal shadow norm. For completeness, and to highlight the importance of finite depth twirling, we show in Fig.~\ref{fig:scaling}, the shadow norm scaling with operator/subregion size $k$ for both the Pauli operators and the second Renyi entropy when evaluated at optimal depth. Even in the presence of noise, the shadow norm scales as an exponential $\sim b^k$ for Pauli operators and $\sim b^{2 \abs{A}}$ for the second Renyi entropy for a region of size $\abs{A}$. In the former case, $b$ goes from $2$ for no noise to $3$ at $f_{\text{th}}$, while it goes from $2.2$ to $2.65$ in the latter.

\paragraph{\textbf{Summary and Discussion.}} Previous works showed that classical shadows based estimation of observables can benefit from twirling with shallow depth unitary circuits, greatly reducing the scaling of sample complexity with  observable size. In this work, we studied how noise affects these results, in addition to computing shadow norms related to estimating Renyi entropies. Working with unitary $n$-designs for $n \ge 2$, we find that any single-qubit noise model can be reduced to an effective depolarizing noise with an effective damping parameter $f$ that can be readily estimated for the relevant experimental realization of qudits. 

As noise increases, the ability of an observer to learn the state $\rho$ via twirling decreases until, at $f= f_{\text{th}}$, it becomes disadvantageous to twirl beyond a single layer of one-qubit scrambling gates, regardless of the size of observable or the region for which the Renyi entropy is being computed; these thresholds are provided in Table~\ref{tab:fvalues}.

For noise below the threshold, twirling up to depth $t_*$ reduces sample complexity before noise and operator spreading overcome the reduction due to bulk relaxation of operators under random unitary dynamics. For finite noise, there is a maximum circuit depth $t_{\text{max}} (f)$ that bounds the optimal depth for estimating any local property of the system. This bound is obtainable using MPS simulations; in practice $t_{\text{max}} = 3$ in even relatively low noise setting with $f = 0.99$, suggesting that the optimization of the sample complexity may always be achieved with small finite depth circuits. A simple mean field analysis supports these findings, but is not accurate enough to estimate the optimal sample complexity. For both the estimation of Pauli operators and the $n^{\text{th}}$ Renyi entropy, the shadow norm evaluated at the minimum $t_*$ scales as $b^k, b^{|A|}$ respectively, where $b$ is a function of $f$, which we provide in Fig. \ref{fig:scaling}.

In this work, we took a noise-agnostic approach and did not try to correct the classical snapshot with knowledge of the noise \cite{KOH2022_ClassicalShadowsWithNoise, JNANE2023_NoisyShadows, SHAO2023_groupShadowNoise, CHEN2021_ShadowEstimation, WU2023_NoisyFermionicShadows}. Alternative post-processing aimed at improving the sample complexity by accounting for noise merit further inquiry; we provide a simple example of such a scheme in the SM. Although specific values of the thresholds and the shadow norm scaling are architecture dependent, qualitative features of this work should hold generally, that is, both noise thresholds $f_{\text{th}}$ and upper bound on the optimal circuit depth $t_{\text{max}}$ should exist and be independent of details of the properties being estimated. It would be insightful to test these thresholds on present-day quantum computers, which are hampered by more general noise and other deficiencies. 

\section{Acknowledgments}
KA and PR acknowledge funding support from NSERC, FRQNT and INTRIQ. This work was also supported by the US Department of Energy, Office of Science, Basic Energy Sciences, Materials Sciences and Engineering Division. NB is supported by the U.S. Department of Energy, Office of Basic Energy Sciences, under Contract No.DE-SC0012704. The authors also acknowledge useful discussions with Matteo Ippoliti and William Coish.

\bibliography{ShadowNoise_V11}

\newpage
\appendix
\begin{widetext}
\addcontentsline{toc}{section}{Appendix} 
\part{Appendix} 
\parttoc

\section{Shift and clock operator}\label{sec:ShiftClockOperator}
The shift and clock operators \cite{GHEORGHIU2014_GeneralizedPaulis} are given respectively by 
\begin{align}
Z = \sum_{n=1}^{q}e^{i2\pi \frac{n}{q}}\ket{n}\bra{n}\nonumber \\
X = \sum_{n=1}^{q}\ket{(n \mod q) + 1}\bra{n}\nonumber
\end{align}
where the states $\ket{n}$ form an orthonormal basis for a Hilbert space of dimension $q$. The operators $\frac{X^nZ^m}{\sqrt{q}}$, $n \in 1,...,q, m \in 1,..,q$ form a basis for all possible linear operators and are orthonormal with respect to the inner product 
$(A,B) \rightarrow \operatorname{Tr}(A^{\dagger}B)$, i.e $\frac{1}{q}\operatorname{Tr}((X^nZ^m)^{\dagger}X^{n'}Z^{m'}) = \delta_{n,n'}\delta_{m,m'}$. Thus, any operator $O$ can be decomposed in terms of the operators $X^nZ^m$ as
\begin{equation}
O = \sum_{n=1,m=1}^q\frac{1}{q}\operatorname{Tr}((X^nZ^m)^{\dagger}O)X^nZ^m.
\end{equation}
The tensor product $\bigotimes_{j=1}^L X_j^{n_j}Z_j^{n_j}$ forms a basis for all linear operators on $L$ qudits of dimension $q$. One may also show that the operator $X$ and $Z$ obey the commutation relations $ZX = e^{\frac{2\pi i}{q}}XZ$. Furthermore, the product $X^nZ^m$ is given by 
\begin{equation}
X^nZ^m=\sum_{j=1}^{q}|(j+n) \text{ mod } q \rangle e^{\frac{2i\pi m j}{q}}\langle j|.
\end{equation}
We note that 
\begin{equation}
\sum_{n,m = 1}^qX^nZ^m = \sum_n q|n \rangle\langle q|
\end{equation}
and similarly, we have that
\begin{equation}\label{eq:maxEigenSum}
\sum_{n_1,m_1,...,n_l,m_l}X_1^{n_1}Z_1^{m_1}\otimes ... \otimes X_l^{n_l}Z_l^{m_l} = \sum_{n_1,...,n_l}q^l\ket{n_1,...,n_l}\bra{q,...,q}.
\end{equation}
Thus, the sum of all Pauli operators with support on $l$ sites produces an operator with a maximum eigenvalue equal to $q^l$, which will be of useful in subsequent sections.

\section{Pauli operators as eigenstates}
\label{app:LocallyScramblingEigenvalues}
We note that
\begin{align}\label{eq:ChannelWithNoise}
\mathcal{M}(\rho) = 
\sum_b \int_{\mathcal{U}(t)} dU(t) \bra{b}K_{\varepsilon,U(t)}(V\rho V^{\dagger})\ket{b} \left(V^{\dagger}U(t)^{\dagger}\ket{b}\bra{b}U(t)V\right) = 
V^{\dagger}\mathcal{M}(V\rho V^{\dagger})V
\end{align}
where $V$ is any product of single-site Clifford operators and we used the fact that the channel $\mathcal{M}$ is locally scrambling. This implies that 
\begin{equation}
V\mathcal{M}(\rho)V^{\dagger} = \mathcal{M}(V\rho V^{\dagger}).
\end{equation}
From there, one may study the effect of the channel on a Pauli operator $P_\beta$. Let us consider the transformation $V_j = Z_j$ for some site $j$. Provided $P_\beta$ at site $j$ is given by $X^{n_{\beta,j}}Z^{m_{\beta,j}}$, one has that 
\begin{align}
Z_j\mathcal{M}(P_\beta)Z_j^{\dagger} = 
\sum_\lambda c_{\beta,\lambda} P_\lambda e^{-i2\pi n_{\lambda,j}/q} = \nonumber \\ 
\sum_\lambda c_{\beta,\lambda} P_\lambda e^{-i2\pi n_{\beta,j}/q} =  \mathcal{M}(Z_j P_\beta Z_j^{\dagger}).
\end{align}
Since the Pauli operators form an orthonormal basis, this is only possible provided $n_{\lambda,j} = n_{\beta,j}$. Repeating the argument with $V = X_j$ yields $m_{\lambda,j} = m_{\beta,j}$ from which we conclude that the only allowed Pauli operator in the expansion of $\mathcal{M}(P_\beta)$ is $P_\beta$ itself, and thus it is an eigenstate of $\mathcal{M}$ regardless of the error channel $\varepsilon$.

\section{Noiseless eigenvalue calculation}\label{app:NoiselessEigenvalueDerivation}
In this section, we compute the eigenvalue $\beta_{i}$ in the noiseless case. First, we note that 

\begin{align}\label{eq:densityChannel}
\mathcal{M}(P_i) = 
 \sum_b\int_{\mathcal{U}(t)} dU(t) \bra{b} V_bU(t)P_i U(t)^\dagger V_b^{\dagger}\ket{b}\left(U(t)^\dagger V_b^{\dagger} \ket{b}\bra{b}V_bU(t)\right) \nonumber \\= D\int_{\mathcal{U}(t)} dU(t) \bra{0} U(t)P_i U(t)^\dagger \ket{0}\left(U(t)^\dagger \ket{0}\bra{0}U(t)\right)
\end{align}

where $D = q^L$ and we have taken advantage of the fact that the integration measure $dU(t)$ is invariant under transformations $U(t) \rightarrow VU(t)$ to introduce the unitary $V_b$  which sends $\ket{b}$ to $\ket{0}$, i.e $V_b^\dagger \ket{b} \rightarrow \ket{0}$, where we define $\ket{0}$ to be the state such that $Z_j\ket{0} = \ket{0}$ for any site $j$. The eigenvalue $\beta_i$ is given explicitly by
\begin{equation}
\beta_i = \frac{1}{D}\operatorname{Tr}(P_{i}^{\dagger}\mathcal{M}(P_i)).
\end{equation}
which can be expanded using Eq. (\ref{eq:densityChannel}) and yields
\begin{equation}\label{eq:eigenvalue}
\beta_i = \int_{\mathcal{U}(t)} dU(t) |\bra{0} U(t)P_i U(t)^\dagger \ket{0}|^2.
\end{equation}
From there, we can expand $U(t)P_iU(t)^\dagger$ in the Pauli basis $U(t)P_iU(t)^\dagger = \sum_\lambda \alpha_{i,\lambda',U(t)} P_{\lambda'}$ for some coefficients $\alpha_{i,\lambda',U(t)}$. By substituting this decomposition in Eq. (\ref{eq:eigenvalue}), one obtains 
\begin{equation}
\beta_i = \sum_{\lambda,\lambda'}\int_{\mathcal{U}(t)} dU(t) \alpha_{i,\lambda,U(t)}^* \alpha_{i,\lambda',U(t)}\bra{0}P_\lambda^{\dagger}\ket{0}\bra{0}P_{\lambda'}\ket{0}.
\end{equation}
Let us now focus on a specific element of the sum and let us assume that $P_\lambda^{\dagger}$ and $P_{\lambda'}$ differ at a site $j$ such that locally $P_\lambda^{\dagger}$ is given by 
$(X_j^{n_{\lambda,j}}Z_j^{m_{\lambda,j}})^{\dagger}$ and $P_{\lambda'}$ is given by $X_j^{n_{\lambda',j}}Z_j^{m_{\lambda',j}}$. 
We may write
\begin{align}
&\int_{\mathcal{U}(t)} dU(t) \alpha_{i,\lambda,U(t)}^*\alpha_{i,\lambda',U(t)}\bra{0}P_\lambda^{\dagger} \ket{0}\bra{0}P_{\lambda'}\ket{0} = \nonumber \\
&\frac{1}{D^2}\int_{\mathcal{U}(t)} dU(t) \operatorname{Tr}\left(P_{\lambda }U(t)P_i^{\dagger}U(t)^{\dagger}\right)\operatorname{Tr}\left(P_{\lambda' }^{\dagger}U(t)P_iU(t)^{\dagger}\right)\bra{0}P_\lambda^{\dagger} \ket{0}\bra{0}P_{\lambda'}\ket{0} = \nonumber \\
&\frac{1}{D^2}\int_{\mathcal{U}(t)} dU(t) \operatorname{Tr}\left(Z_j^{\dagger}P_{\lambda }Z_jU(t)P_i^{\dagger}U(t)^{\dagger}\right)\operatorname{Tr}\left(Z_j^{\dagger}P_{\lambda' }^{\dagger}Z_jU(t)P_iU(t)^{\dagger}\right)\bra{0}P_\lambda^{\dagger} \ket{0}\bra{0}P_{\lambda'}\ket{0} = 
\nonumber \\
&\left(\int_{\mathcal{U}(t)} dU(t) \alpha_{i,\lambda,U(t)}^*\alpha_{i,\lambda',U(t)}\bra{0}P_\lambda^{\dagger} \ket{0}\bra{0}P_{\lambda'}\ket{0}\right)e^{2\pi i(n_{\lambda,j} - n_{\lambda',j})/q}
\end{align}
where we used the commutation relation $XZ = ZXe^{2\pi i/q}$ and the locally scrambling property to send $U(t) \rightarrow Z_jU(t)$, where $Z_j$ is the $Z$ Pauli operator acting on site $j$. 
This is obviously only possible if $n_{\lambda,j} = n_{\lambda',j}$. A similar argument can be used to show that $m_{\lambda,j} = m_{\lambda',j}$ from which we conclude that $\lambda$ must be equal to $\lambda'$. Consequently, we get 
\begin{align}
\beta_i =  \sum_\lambda\left(\int_{\mathcal{U}(t)} dU(t) |\alpha_{i,\lambda,U(t)}|^2\right) |\bra{0}P_\lambda \ket{0}|^2.
\end{align}
Denoting $\int_{\mathcal{U}(t)} dU(t) |\alpha_{i,\lambda,U(t)}|^2$ by $\overline{|\alpha_{i,\lambda,U(t)}|^2}$, we have 
\begin{align}
\beta_i = \sum_\lambda\overline{|\alpha_{i,\lambda,U(t)}|^2} |\bra{0}P_\lambda \ket{0}|^2.
\end{align}
Due to the locally scrambling property, the average $\overline{|\alpha_{i,\lambda,U(t)}|^2} $ can only depend on the specific weight distribution $w_j$ of the Pauli operator, where $w_j = 0$ if the Pauli operator is identity at site $j$, and $w_j = 1$ otherwise. We may thus replace the $\lambda$ dependency in $\overline{|\alpha_{i,\lambda,U(t)}|^2} $ by a dependency on $\vec{w}$, the vector of weights. Doing so yields
\begin{equation}
\beta_i = \sum_{\vec{w}}\sum_{\lambda:\vec{w}_\lambda = \vec{w}}\overline{|\alpha_{i,\lambda,U(t)}|^2} |\bra{0}P_\lambda \ket{0}|^2,
\end{equation}
where $\vec{w}_\lambda$ is the weight vector of the Pauli operator $P_\lambda$.
For a given weight configuration $\vec{w}$, one has that $|\bra{0}P_\lambda \ket{0}|^2$ is non-zero and equal to $1$ only if $P_\lambda$ is a string of $Z$ operators.
There are $(q - 1)^{\norm{\vec{w}}^2}$ Pauli operators that are strings of $Z$ Pauli operators for a fixed weight vector $\vec{w}$ out of $(q^2 - 1)^{\norm{\vec{w}}^2}$ possible Pauli operators. Using this, we have that
\begin{align}
\beta_i = \sum_{\vec{w}}\left(\sum_{\lambda:\vec{w}_\lambda = \vec{w}}\overline{|\alpha_{i,\lambda,U(t)}|^2}\right)\frac{(q - 1)^{\norm{\vec{w}}^2}}{(q^2 - 1)^{\norm{\vec{w}}^2}} = 
\sum_{\vec{w}}\left(\sum_{\lambda:\vec{w}_\lambda = \vec{w}}\overline{|\alpha_{i,\lambda,U(t)}|^2}\right)\frac{1}{(q + 1)^{\norm{\vec{w}}^2}} 
\end{align}
The factor $\left(\sum_{\lambda: \vec{w}_\lambda = \vec{w}} \overline{|\alpha_{i,\lambda,U(t)}|^2} \right)$ is the probability $\text{Pr}(\vec{w})$ that the initial operator $P_i$ has weight $\vec{w}$ after the application of $U(t)$. We may thus write 
\begin{equation}
\beta_i = \sum_{\vec{w}}\text{Pr}(\vec{w})\frac{1}{(q + 1)^{\norm{\vec{w}}^2}}. 
\end{equation}

\section{Noisy eigenvalue calculation}
The eigenvalues of the noisy channel $\mathcal{M}$ may be computed as
\begin{align}\label{eq:ShadowNorm}
\beta_{i,\varepsilon} = \frac{1}{D}\operatorname{Tr}(P_{i}^{\dagger}\mathcal{M}(P_{i}))
= \int_{\mathcal{U}(t)} dU(t)\bra{0}K_{\varepsilon,U(t)}(P_{i}) \ket{0}\bra{0} U(t)P_{i}^{\dagger} U(t)^\dagger \ket{0}.
\end{align} Now, we may decompose $K_{\varepsilon,U(t)}(P_{i})$ in the Pauli basis to get $K_{\varepsilon,U(t)}(P_{i}) = \sum_\lambda \alpha_{i,\lambda,U(t),\varepsilon}P_{\lambda}$ and 
$U(t)P_{i}^{\dagger}U(t)^\dagger = \sum_\lambda \alpha_{i,\lambda,U(t)}^*P_{\lambda}^{\dagger}$ . Following closely the calculations for the noiseless case, one obtains 
\begin{align}\label{eq:NoisyEigenvalueRaw}
\beta_{i,\varepsilon} = \sum_{\vec{w}}\left(\sum_{\lambda: \vec{w}_\lambda = \vec{w}} \overline{\alpha^*_{i,\lambda,U(t)}\alpha_{i,\lambda,U(t),\varepsilon}} \right)\frac{1}{(q + 1)^{\norm{\vec{w}}^2}}
\end{align}
where 
\begin{align}
\overline{\alpha^*_{i,\lambda,U(t)}\alpha_{i,\lambda,U(t),\varepsilon}}  = \int dU(t)\alpha^*_{i,\lambda,U(t)}\alpha_{i,\lambda,U(t),\varepsilon}.
\end{align}
Taking advantage of the fact that the multi-qudits Clifford group is a 2-design \cite{WEBB2016_Qudit2Design}, we compute the average in Eq. (\ref{eq:NoisyEigenvalueRaw}) with Clifford unitaries

\begin{align}
\beta_{i,\varepsilon} =  \sum_{w}\left(\sum_{\lambda: \norm{\vec{w}_\lambda}^2 = w} \frac{1}{|\mathcal{C}(t)|}\sum_{U(t) \in \mathcal{C}(t)}\alpha^*_{i,\lambda,U(t)}\alpha_{i,\lambda,U(t),\varepsilon} \right)\frac{1}{(q + 1)^{w}}
\end{align}
where $\mathcal{C}(t)$ is the set of all circuits of depth $t$ where the local unitaries acting on $n$ qudits of dimension $q$ are selected at random from $\mathcal{C}_n^q$, the Clifford group on $n$ qudits.  We have denoted above the total number of elements in $\mathcal{C}(t)$ by $|\mathcal{C}(t)|$. For the rest of this section, we will assume that the error channel $\varepsilon$ acts as follows on the Pauli operators 
\begin{align}
\varepsilon: X^{n}Z^m \rightarrow f_{n,m}X^nZ^m
\end{align}
where $f_{q,q} = 1$. We show in the next section that this actually encapsulates all possible single-site error channels $\varepsilon$. We have that Clifford gates send Pauli operators to Pauli operators, and in this setting, the Pauli operators are eigenstates of the error channel $\varepsilon$. This implies that we may view the evolution of the Pauli operator $P_{i}$ under $K_{\varepsilon, U(t)}$ as a sequence of Pauli operators with distinct weight configurations (we obtain a Pauli operator with a possibly distinct weight configuration after the application of each layer composed of two-site Clifford gates).
For a given sequence of operators $\{\lambda_0,...,\lambda_{t+1}\}$ where $\lambda_0$ is the Pauli operator before the application of the first unitary layer and $\lambda_{t+1}$ the Pauli operator after the application of that last unitary layer, we denote by $w_{\text{tot},(n,m)}$ the number of times the operator $X^nZ^m$ appears in the sequence $\{\lambda_0,...,\lambda_t\}$ where $\lambda_{t+1}$ is omitted, which accounts for the fact the the error channel is only applied $t+1$ times. Then, the coefficients $\alpha_{i,\lambda,U(t),\varepsilon}$ satisfy the relation $\alpha_{i,\lambda,U(t),\varepsilon} = \alpha_{i,\lambda,U(t)}\prod_{(n,m)}f_{(n,m)}^{w_{\text{tot},(n,m)}}$. Since the full circuit sends Pauli operators to Pauli operators, we get that $\alpha_{i,\lambda,U(t)}\alpha_{i,\lambda,U(t)}^*$ must be either $1$ or $0$. Thus, we can replace the sum over Clifford unitary gates by a sum over sequences of Pauli operators $\{\lambda_0, ..., \lambda_{t+1}\}$ weighted by the number of times $N_i(\lambda_0,...,\lambda_{t+1})$ that 
given sequence occurs. Thus we may write

\begin{align}
\beta_{i,\varepsilon} = \sum_{w}\left(\sum_{\lambda_t: \norm{\vec{w}_{\lambda_t}}^2 = w} \sum_{\lambda_0,...,\lambda_{t-1},\lambda_{t+1}} \frac{N_i(\lambda_0,...,\lambda_{t+1} )\prod_{(n,m)}f_{(n,m)}^{w_{\text{tot},(n,m)}}}{|\mathcal{C}(t)|} \right)\frac{1}{(q + 1)^{w}} \nonumber =  \\ 
\sum_{w}\left(\sum_{\lambda_t: \norm{\vec{w}_{\lambda_t}} = w} \sum_{\lambda_0,...,\lambda_{t-1},\lambda_{t+1}} \text{Pr}(\lambda_0,...,\lambda_{t+1})\prod_{(n,m)}f_{(n,m)}^{w_{\text{tot},(n,m)}} \right)\frac{1}{(q + 1)^{w}} = \nonumber \\
\sum_{\vec{w}(t)}\left(\sum_{\vec{w}(1),...,\vec{w}(t-1)}\text{Pr}(\vec{w}(0),...,\vec{w}(t))\left(\sum_{(n,m)\neq (0,0)}\frac{f_{(n,m)}}{(q^2 - 1)}\right)^{w_{\text{tot}}}\right)\frac{1}{(q+1)^{\norm{\vec{w}(t)}^2}}
\end{align}

where we interpreted the ratio $\frac{N_i(\lambda_0,...,\lambda_{t+1})}{|\mathcal{C}(t)|}$ as a probability $\text{Pr}(\lambda_0,...,\lambda_{t+1})$. Since there is a total of $(q^2-1)^{\norm{\vec{w}(t+1)}^2}(q^2 - 1)^{w_{\text{tot}}}$ sequences with weights $\vec{w}(0),..., \vec{w}(t+1)$, 
where $w_{\text{tot}} = \sum_{j=0}^{t}\norm{\vec{w}(j)}^2$, we get that  $\text{Pr}(\lambda_0,...,\lambda_{t+1}) = \text{Pr}(\vec{w}(0),...,\vec{w}(t))\frac{1}{(q^2 - 1)^{w_{\text{tot}}}(q^2-1)^{\norm{\vec{w}(t+1)}^2}}$ provided
$\vec{w}_{\lambda_j} = \vec{w}(j)$ for $j = 0,1,2,...,t+1$ and we used the fact that the probability
$\text{Pr}(\lambda_0,...,\lambda_{t+1})$ only depends on the weight configurations. The weight vector $\vec{w}(t+1)$ is always equal to $\vec{w}(t)$, and the noise factor is independent of $\vec{w}(t+1)$, implying that the sum over $\lambda_{t+1}$ removes the factor $\frac{1}{(q^2-1)^{\norm{\vec{w}(t+1)}^2}}$ and leads to Eq. (\ref{eq:numericalEq}). The final equation shows that the case where the depolarizing strengths $f_{(n,m)}$ are not equal can be studied via a uniform depolarizing channel where the noise parameter $f$ is given by the average $f \equiv \sum_{(n,m)\neq (0,0)}\frac{f_{(n,m)}}{q^2-1}$, which was assumed throughout this work.

\section{Inconsequential noise parameters}\label{app:f0Parameter}
Consider the eigenvalue equation 
\begin{align}\label{eq:eigenvalueEq}
\beta_{i,\varepsilon} = \frac{1}{D}\operatorname{Tr}(P_i^{\dagger}\mathcal{M}(P_i))  
= \int_{\mathcal{U}(t)} dU(t)\bra{0}K_{\varepsilon, U(t)}(P_i) \ket{0}\bra{0} U(t)P_i^{\dagger} U(t)^\dagger \ket{0}.
\end{align} In what follows, we will make use of the following known result \cite{DANKERT2009_HaarIntegral}
\begin{align}\label{eq:HaarIntegral}
\frac{1}{|\mathcal{C}_{2}^q|}\sum_{h_{j,j+1} \in \mathcal{C}_{2}^q} h_{j,j+1}^{\dagger} A h_{j,j+1} B h_{j,j+1}^{\dagger} C h_{j,j+1}=\frac{\operatorname{Tr}(A C) \operatorname{Tr}(B)}{q^2} \frac{\mathbb{1}}{q^2}+\nonumber \\ \left(\frac{q^2 \operatorname{Tr}(A) \operatorname{Tr}(C)-\operatorname{Tr}(A C)}{q^2\left(q^{4}-1\right)}\right)\left(B-\operatorname{Tr}(B) \frac{\mathbb{1}}{q^2}\right),
\end{align} where $\mathcal{C}_{n}^q$ denotes the Clifford group on $n$ qudits of dimension $q$, and $|\mathcal{C}_{n}^q|$ is the number of elements in the group. 
From here, we replace the average in Eq. (\ref{eq:eigenvalueEq}) with an average over the multi-qudit Clifford group, which is equivalent as it forms a two-design \cite{WEBB2016_Qudit2Design}.
Defining the error channel acting on a full layer $\otimes_{j=1}^L\varepsilon_j \equiv \varepsilon_L$, we consider its Krauss representation $\varepsilon_L(A) : A \rightarrow \sum_jK_jAJ_j^{\dagger}$, which we use to write Eq. (\ref{eq:eigenvalueEq}) as 

\begin{align}
&\frac{1}{|\mathcal{C}(t)|}\sum_{U(t)\in \mathcal{C}(t)}\bra{0}K_{\varepsilon,U(t)}(P_i) \ket{0}\bra{0} U(t)P_i^{\dagger} U(t)^\dagger \ket{0} = \nonumber \\ 
&\frac{1}{|\mathcal{C}(t)|}\sum_{U_0\in \mathcal{C}_0, U_1 \in \mathcal{C}_1,...,U_{0}' \in \mathcal{C}_{0}'}\bra{0}\sum_{j_0,j_1,j_2,...,j_t}U_0'K_{j_t}U_t K_{j_{t-1}}...K_{j_1}U_1K_{j_0}U_0P_iU_0^{\dagger}J_{j_0}U_1^{\dagger}J_{j_1}^{\dagger}...U_{t-1}^{\dagger}J_{j_{t-1}}^{\dagger}U_t^{\dagger}J_{j_{t}}^{\dagger}U_0'^{\dagger} \ket{0}\times \nonumber \\ &\bra{0} U_0'U_t...U_1U_0P_i^{\dagger} U_0^{\dagger}U_1^{\dagger}...U_t^{\dagger}U_0'^{\dagger} \ket{0}
\end{align}
where $\mathcal{C}_l$ is the set of unitaries where each unitary in the $l^{\text{th}}$ layer is chosen at random from the corresponding Clifford group on two or one qudit.
We will show that the error channel $\varepsilon_L$ can be replaced by a channel where the individual terms $f_{(n_j,m_j),(n_j',m_j')}$ of the individual single-site channels $\varepsilon_j$ that compose $\varepsilon_L$ have been set to $0$ for all pairs $(n_j,m_j) \neq (n_j',m_j')$. Consider the following term in the above sum 
\begin{align}\label{eq:ExplicitCalculationf0}
&\sum_{U_1 \in \mathcal{C}_1} U_1
\left(\sum_{j_0}K_{j_0}U_0P_iU_0^{\dagger}J_{j_0}^{\dagger}\right)U_1^{\dagger}\left(J_{j_1}U_2^{\dagger}J_{j_2}^{\dagger}U_3^{\dagger}...U_{t-1}^{\dagger}J_{j_{t-1}}^{\dagger}U_t^{\dagger}J_{j_{t}}^{\dagger}U_0'^{\dagger}\ket{0}\bra{0} U_0'U_t...U_3  U_2\right)U_1\left(U_0P_i^{\dagger}U_0^{\dagger}\right)U_1^{\dagger} = \nonumber \\
&\sum_{U_1 \in \mathcal{C}_1} U_1\left(\varepsilon_L(U_0P_iU_0^{\dagger})\right)U_1^{\dagger}RU_1\left(U_0P_i^{\dagger}U_0^{\dagger}\right)U_1^{\dagger}
\end{align}
where we have defined $R = J_{j_1}^{\dagger}U_2^{\dagger}...U_{t-1}^{\dagger}J_{j_{t-1}}^{\dagger}U_t^{\dagger}J_{j_{t}}^{\dagger}U_0'^{\dagger} \ket{0}\bra{0} U_0'U_t..U_2$. We see that $\varepsilon_L(U_0P_iU_0^{\dagger})$ can be thought of as a superposition of Pauli operators where some sites which originally hosted the Pauli operator $X_{2s-1}^{n_{2s-1}}Z_{2s-1}^{m_{2s-1}}X_{2s}^{n_{2s}}Z_{2s}^{m_{2s}}$ at the sites $S = \{2s-1,2s\}$ are replaced by $f_{(n_{2s},m_{2s}),(n'_{2s}, m'_{2s})}f_{(n_{2s-1},m_{2s-1}),(n'_{2s-1}, m'_{2s-1})}X_{2s-1}^{n'_{2s-1}}Z_{2s-1}^{m'_{2s-1}}X_{2s}^{n'_{2s}}Z_{2s}^{m'_{2s}}$.
Considering one such operator, we denote the Pauli operator on the remaining sites by $P_{\varepsilon_L,U,S^c}$  and we denote by $P_{U,S^c}$ the Pauli operator on the remaining sites of $U_0P_iU_0^{\dagger}$. Without loss of generality, we focus on a pair of sites $2s-1, 2s$ that have the same support as one of the Haar-random unitary gates $h_{2s-1,2s}$ in the layer $U_1$. 
Finally, let us write down the operator $R$ as a sum of Pauli operators, where we separate the part acting on the sites $S = \{2s-1,2s\}$ and the rest of the system, i.e $R = \sum_{\lambda}c_{\lambda}P_{\lambda,S}P_{\lambda,S^{c}}$. Focusing on one of the Pauli operators that appear in a decomposition of $\varepsilon_L(U_0P_iU_0^{\dagger})$ with local Pauli operators replaced at sites $2s-1,2s$, we get that Eq. (\ref{eq:ExplicitCalculationf0}) for that particular term is proportional to 
\begin{align} 
&\sum_{\lambda}c_\lambda \sum_{h_{2j-1,2j} \in \mathcal{C}_{2}^q, j \neq s}\left(\prod_{j\neq s}h_{2j-1.2j}\right)P_{\varepsilon_L,U,S^c}\left(\prod_{j\neq s}h_{2j-1,2j}^{\dagger}\right) P_{\lambda,S^c} \left(\prod_{j\neq s}h_{2j-1.2j}\right)P_{U,S^c}^{\dagger}\left(\prod_{j\neq s}h_{2j-1,2j}^{\dagger}\right) \times \nonumber \\ 
&\left(\sum_{h_{2s-1,2s} \in \mathcal{C}_{2}^q}h_{2s-1,2s}f_{(n_{2s-1},m_{2s-1}),(n'_{2s-1}, m'_{2s-1})}f_{(n_{2s},m_{2s}),(n'_{2s}, m'_{2s})}X_{2s-1}^{n'_{2s-1}}Z_{2s-1}^{m'_{2s-1}}X_{2s}^{n'_{2s}}Z_{2s}^{m'_{2s}}\times\right. \nonumber \\ 
&\left. h_{2s-1,2s}^{\dagger}P_{\lambda,S}h_{2s-1,2s}(Z_{2s-1}^{\dagger})^{n_{2s-1}}(X_{2s-1}^{\dagger})^{m_{2s-1}}(X_{2s}^{\dagger})^{n_{2s}}(Z_{2s}^{\dagger})^{m_{2s}}\right).
\end{align} Now the sum over $h_{2s-1,2s}$ above can be carried over with the identification $A = X_{2s-1}^{n'_{2s-1}}Z_{2s-1}^{m'_{2s-1}}X_{2s}^{n'_{2s}}Z_{2s}^{m'_{2s}}, B = P_{\lambda,S}, C = (Z_{2s-1}^{\dagger})^{n_{2s-1}}(X_{2s-1}^{\dagger})^{m_{2s-1}}(X_{2s}^{\dagger})^{n_{2s}}(Z_{2s}^{\dagger})^{m_{2s}}$. Since $\operatorname{Tr}(A)\operatorname{Tr}(C) = 0, \operatorname{Tr}(AC) = 0$ for all cases where $(n_{2s-1},m_{2s-1})\neq (n'_{2s-1},m'_{2s-1})$ or $(n_{2s},m_{2s})\neq (n'_{2s},m'_{2s})$, these terms do not contribute. Thus, the factors $f_{(n_j,m_j),(n_j',m_j')}, (n_j,m_j) \neq (n_j',m_j')$ in the first application of the error channel may all be taken to $0$ without changing the value of the integral. After doing so, the Pauli operators become eigenstates of the error channel applied after the first unitary layer $U_0$ and only acquire a multiplicative factor. This implies that a similar argument can be carried over sequentially for each noise layer, which completes the proof that the factors $f_{(n_j,m_j),(n_j',m_j')}, (n_j,m_j) \neq (n_j',m_j')$ may be taken to be $0$ without changing the value of the integral.

\section{Noisy shadow norm calculation}\label{app:NoisyShadowNorm}
The shadow norm is an upper bound on the variance of the estimators $o_{U(t),b}$. Due to local scrambling, the variance turns out to be be independent of $\rho$ and is thus equal to the shadow norm. Explicitly, we have 
\begin{equation}
\operatorname{Var}(o_{U(t),b}) = \mathbb{E}_{U(t),b}(|o_{U(t),b}|^2) - |\mathbb{E}_{U(t),b}(o_{U(t),b})|^2.
\end{equation}
Now $|\mathbb{E}_{U(t),b}(o_{U(t),b})|^2 = |\operatorname{Tr}(\rho P_{i})|^2 \approx O(1)$ for any Pauli operator $P_{i}$. Thus, this term can be neglected in front of  $\mathbb{E}_{U(t),b}(|o_{U(t),b}|^2)$, which as we show grows exponentially with the weight of $P_{i}$. We have that 
\begin{align}
\mathbb{E}_{U(t), b}\left[|o_{U(t),b}|^2\right]=  \int_{\mathcal{U}(t)} \mathrm{d} U(t) \sum_b\bra{b}K_{\varepsilon,U(t)}(\rho)\ket{b}|\operatorname{Tr}[P_{i} \mathcal{M}^{-1}(U(t)^{\dagger}\ket{b}\bra{b} U(t))]|^2.
\end{align}
Using the fact that the set $\mathcal{U}(t)$ is locally scrambling, we can reduce this expression to 
\begin{align}
\mathbb{E}_{U(t), b}\left[|o_{U(t),b}|^2\right]=  D\int_{\mathcal{U}(t)} \mathrm{d} U(t) \bra{0}K_{\varepsilon,U(t)}(\rho)\ket{0}|\operatorname{Tr}[P_{i} \mathcal{M}^{-1}(U(t)^{\dagger}\ket{0}\bra{0} U(t))]|^2.
\end{align} Now, expanding $U(t)^{\dagger}\ket{0}\bra{0} U(t)$ in the Pauli basis as $\sum_\lambda c_{\lambda,U(t)}P^{\dagger}_{\lambda}$ leads to 
\begin{align}
\operatorname{Tr}[P_{i} \mathcal{M}^{-1}(U(t)^{\dagger}\ket{0}\bra{0} U(t))] = 
\frac{c_{i,U(t)}}{\beta_{i,\varepsilon}}D
\end{align}
where $\beta_{i,\varepsilon}$ is the eigenvalue of the Pauli operator $P_i$ in the presence of noise and we used the fact that both $P_i$ and $P_i^{\dagger}$ have the same eigenvalue.
The coefficient $c_{i, U(t)}$ is given by 
\begin{align}
\frac{1}{D}\operatorname{Tr}[P_{i}U(t)^{\dagger}\ket{0}\bra{0}U(t)] = \frac{1}{D}\bra{0}U(t)P_{i}U(t)^{\dagger}\ket{0}
\end{align}
where we used the cyclicity of the trace. Thus, we obtained  
\begin{align}\label{eq:shadow_norm_noise_before_final}
\mathbb{E}_{U(t), b}\left[|o_{U(t),b}|^2\right]= \frac{D}{(\beta_{i,\varepsilon})^2}\int_{\mathcal{U}(t)} \mathrm{d}U(t) \bra{0}K_{\varepsilon,U(t)}(\rho)\ket{0}|\bra{0}U(t)P_{i}U(t)^{\dagger}\ket{0}|^2.
\end{align}
Now, due to the locally scrambling property of  $\mathcal{U}(t)$, any non-zero weight Pauli operator that appears in a decomposition of $\rho$  must vanish, as otherwise a unitary $V$ can be introduced to add a phase in front of the integral. Thus, the only contributing term in a decomposition of $\rho$ is $I/D$, which leads to 
\begin{align}\label{eq:shadow_norm_noise_final}
\mathbb{E}_{U(t), b}\left[|o_{U(t),b}|^2\right]= \frac{1}{(\beta_{i,\varepsilon})^2}\int_{\mathcal{U}(t)} \mathrm{d}U(t) |\bra{0}U(t)P_{i}U(t)^{\dagger}\ket{0}|^2.
\end{align}
We note that this integral corresponds to the noiseless eigenvalue $\beta_{i}$, leaving us with 
\begin{align}
&\mathbb{E}_{U(t), b}\left[|o_{U(t),b}|^2\right]= \frac{\beta_{i}}{(\beta_{i,\varepsilon})^2}
\end{align}
and consequently 
\begin{align}
\norm{P_i}^2_{\text{sh}} = \frac{\beta_i}{\beta_{i,\varepsilon}^2} 
\end{align}

\section{Lower bound on noise thresholds}
In this section, we discuss the derivation of the lower bound on the thresholds $f_{\text{th}}$. For a given Pauli operator $P_i$, the first layer of the circuit will act on $k_1$ pairs of qudits that contain a single occupied site and will act on $k_2$ pairs of qudits that contain two occupied sites. The total initial weight of the Pauli operator is then given by $\norm{\vec{w}_i}^2 = k_1 + 2k_2$. The value of $f_{\text{th}}$ is found by computing exactly the shadow norm after the application of the first layer. Since the Clifford group sends any Pauli operator to any other Pauli operator with equal frequency, we have that each site becomes doubly occupied with probability $1-2a$, and becomes singly occupied with probability $2a$ where $a = 1/(q^2+1)$. Thus, the noisy eigenvalue at $t = 1$ for a Pauli operator $P_{i}$ with a total weight $\norm{\vec{w}_i}^2 = k_1 + 2k_2$ is given by 
\begin{align}
&\beta_{i,\varepsilon}|_{t=1} =
 f^{k_1 + 2k_2}\sum_{n=0}^{(k_1 + k_2)}\binom{(k_1 + k_2)}{n}\left(\frac{f}{q+1}\right)^{n + (k_1 + k_2)} (1-2a)^n(2a)^{(k_1 + k_2) - n} \nonumber \\ &= f^{k_1 + 2k_2}\left(\left(\frac{f}{q+1}\right)^2(1-2a) + 2a\left(\frac{f}{q+1}\right)\right)^{k_1 + k_2}. 
\end{align}
To obtain the noiseless eigenvalue $\beta_{i}$, we simply take the limit $f = 1$ which yields the shadow norm
\begin{align}\label{eq:ShadowNormAtT=1}
\norm{P_{i}}^2_{\text{sh}}|_{t=1} =  
\frac{1}{f^{2(k_1 + 2k_2)}}\left(\frac{\left(\frac{1}{q+1}\right)^2(1-2a) + 2a\left(\frac{1}{q+1}\right)}{\left(\left(\frac{f}{q+1}\right)^2(1-2a) + 2a\left(\frac{f}{q+1}\right)\right)^2}\right)^{(k_1 + k_2)}.
\end{align}
It is only beneficial to twirl for $t \geq 1$ provided the shadow norm at $t = 1$ is smaller than $(q+1)^{(k_1 + 2k_2)}$, the $t = 0$ shadow norm. The lowest value of $f_{\text{th}}$ is found in the limit $k_1 \rightarrow 0$ which maximises the effect of bulk decay as this maximises the initial Pauli density. In this setting, putting Eq. (\ref{eq:ShadowNormAtT=1}) equal to $(q+1)^{(k_1 + 2k_2)}$ leads to the polynomial equation 

\begin{equation}
f_{\text{th}}^6 (f_{\text{th}} (q-1)+2)^2 - (q^2+1) = 0.
\end{equation}
whose solutions for distinct values of $q$ are displayed in the main text. Since this threshold is independent of $k$, it holds for any Pauli operator. Furthermore, as discussed in the main text, the shadow norm of any observable is given by an appropriate sum of products of the shadow norm of individual Pauli operators, which implies that the above lower bound applies to any observable $O$.

\section{SWAP operator on qudits}\label{app:SWAP}
In this section, we show that the SWAP operator on a pair of qudits $1,2$ with local Hilbert space dimension $q$ can be written down in terms of generalized Pauli operators as 
\begin{align}
\text{SWAP}_{1,2} = \sum_{n=1,m=1}^q\frac{1}{q}e^{\frac{2i\pi(nm)}{q}}X_1^{n}Z_1^{m}\otimes X_2^{-n}Z_2^{-m}
\end{align} where the lower indices on the operators $X,Z$ are only introduced for clarity, i.e $X_j = X$, $Z_j = Z$ $\forall \quad j$.
To prove that this is the right generalized Pauli operator representation of the SWAP operator, we show that it satisfies the relation 
\begin{align}\label{eq:SWAPCondition}
\text{SWAP}_{1,2}(X_1^{i}Z_1^{j}\otimes X_2^{k}Z_2^{l})\text{SWAP}_{1,2}^{\dagger} =  X_2^{k}Z_2^{l}\otimes X_1^{i}Z_1^{j}.
\end{align}
Replacing our expression for the SWAP operator in the l.h.s of Eq. (\ref{eq:SWAPCondition}) yields 

\begin{align}
\sum_{n=1,m=1,n'=1,m'=1}^{q}\frac{e^{\frac{2\pi i(nm - n'm')}{q}}}{q^2}X_1^{(n - n') + i}Z_1^{(m - m') + j}\otimes X_2^{(n'- n) + k}Z_2^{(m' - m) + l}e^{2i\pi\frac{(-n'(j-m') + m(i-n') + n'(l+m') - m(k+n'))}{q}}.
\end{align}  
Now we perform the change of variables $u = n' - n, v = m' - m, n = n, m = m$, which yields 
\begin{align}
&\sum_{n=1}^q\sum_{m=1}^q\sum_{u=1-n}^{q-n}\sum_{v=1-m}^{q-m}\frac{1}{q^2}e^{2i\pi\frac{u((l-j) + v)}{q}}\left( e^{\frac{2\pi i(m((i-k)-u))}{q}}  e^{-\frac{2\pi i(n((j-l) - v))}{q}} \right)X_1^{i-u}Z_1^{j-v}\otimes X_2^{u + k}Z_2^{v + l} = \nonumber \\
&\sum_{n=1}^q\sum_{m=1}^q\sum_{u=1}^{q}\sum_{v=1}^{q}\frac{1}{q^2}e^{2i\pi\frac{u((l-j) + v)}{q}}\left( e^{\frac{2\pi i(m((i-k)-u))}{q}}  e^{-\frac{2\pi i(n((j-l) - v))}{q}} \right)X_1^{i-u}Z_1^{j-v}\otimes X_2^{u + k}Z_2^{v + l} = \nonumber \\ 
&X_1^{k}Z_1^{l}\otimes X_2^{i}Z_2^{j}
\end{align} 
where we have used the fact that the integers $u,v$ appearing either in the exponential factor or as powers of the generalized Pauli matrices can be taken modulo $q$ without changing the result to reorganize the sum over $u,v$ and in the second line we have used the fact that the sum over $m$, $n$ vanishes unless $u = (i-k), v = (j-l)$, which when substituted directly yields the desired result. We now also provide an expression for the operator $S_{n}$ which acts as $S_{n}\ket{i_1,i_2,...,i_n} = \ket{i_n,i_1,...,i_{n-1}}$ on computational basis states, and is obtained by taking the product $S_{n} = \text{SWAP}_{1,2}\text{SWAP}_{2,3}...\text{SWAP}_{n-1,n}$. We get 
\begin{align}
&S_n = \sum_{n_1,m_1,n_2,m_2,...,n_{n-1},m_{n-1}}\frac{1}{q^{n-1}}e^{\frac{2i\pi(\sum_{j=1}^{n-1} n_jm_j)}{q}}X_{1}^{n_1}Z_1^{m_1}\nonumber \\ &\otimes X_{2}^{-n_1}Z_2^{-m_1}X_{2}^{n_2}Z_2^{m_2}\otimes ...X_{n-1}^{-n_{n-2}}Z_{n-1}^{-m_{n-2}}X_{n-1}^{n_{n-1}}Z_{n-1}^{m_{n-1}}\otimes X_n^{-n_{n-1}}Z_n^{-m_{n-1}}.
\end{align}
Reorganizing, this yields 
\begin{align}\label{eq:SwapCoefficients}
&S_n =  
\sum_{n_1,m_1,n_2,m_2,...,n_{n-1},m_{n-1}}\frac{1}{q^{n-1}}e^{\frac{2i\pi(\sum_{j=1}^{n-2} m_j(n_{j} - n_{j+1}) + n_{n-1}m_{n-1})}{q}}X_{1}^{n_1}Z_1^{m_1}\nonumber \\ &\otimes X_{2}^{n_2 - n_1}Z_2^{m_2 - m_1} \otimes ...X_{n-1}^{n_{n-1} - n_{n-2}}Z_{n-1}^{m_{n-1} - m_{n-2}} \otimes X_n^{-n_{n-1}}Z_n^{-m_{n-1}}.
\end{align}

\section{Renyi entropy in terms of the SWAP operator}
In this section we show that $\operatorname{Tr}\left(\rho_A^n\right)$ may be computed as $\operatorname{Tr}\left(\left(S_{A,n}\right)\bigotimes_{l=1}^{n}\rho_l \right)$ where $S_{A,n}$ is the subsystem SWAP operator which acts as $S_{A,n}\ket{i_1,i_2,...,i_n}_A\otimes\ket{j_1,...,j_n}_{A^c} = \ket{i_n,i_1,...,i_{n-1}}_A\otimes\ket{j_1,...,j_n}_{A^c}$ on computational basis states, with $A^c$ the complement of $A$. First, we decompose the copies $\rho_j$ as
\begin{equation}
\rho_j = \sum_{i_j,k_j,l_j,m_j}c_{i_j,k_j,l_l,m_j}\ket{i_j}_A\ket{k_j}_{A^c}\bra{l_j}_A\bra{m_j}_{A^c}.
\end{equation}
The tensor product of the copies $\rho_j$ then yields 

\begin{align}
\bigotimes_{l=1}^{n}\rho_l = 
\sum_{i_1,k_1,l_1,m_1,...,i_{n},k_{n},l_{n},m_{n}}\left(\prod_{j=1}^{n}c_{i_j,k_j,l_j,m_j}\right)
\left(\bigotimes_{s=1}^{n}\ket{i_{s}}_A\ket{k_{s}}_{A^c}\bra{l_{s}}_A\bra{m_{s}}_{A^c}\right).
\end{align}
Once we apply the operator $S_{A,n}$ on the tensor product, we get 
\begin{align}
&S_{A,n}\bigotimes_{l=1}^{n}\rho_l = \sum_{i_1,k_1,l_1,m_1,...,i_{n},k_{n},l_{n},m_{n}}\left(\prod_{j=1}^{n}c_{i_j,k_j,l_j,m_j}\right)
\left(\bigotimes_{s=1}^{n}\ket{i_{s+1}}_A\ket{k_{s}}_{A^c}\bra{l_{s}}_A\bra{m_{s}}_{A^c}\right)
\end{align}
where $i_{n+1} \equiv i_{1}$. Now taking the trace yields 
\begin{align}
&\operatorname{Tr}\left(S_{A,n}\bigotimes_{l=1}^{n}\rho_l\right) = \sum_{i_1,k_1,...,i_{n},k_{n}}\left(\prod_{j=1}^{n}c_{i_j,k_j,i_{j+1},k_j}\right).
\end{align}
Now let us compute $(\rho_A)_j$, which yiedls 
\begin{align}
(\rho_A)_j = \operatorname{Tr}_{A^c}\left(\rho_j\right) = 
\sum_{i_j,k_j,l_j}c_{i_j,k_j,l_j,k_j}
\ket{i_{j}}_A\bra{l_{j}}_A.
\end{align}
The product $((\rho_A)_j)^n$ is given by 
\begin{equation}
((\rho_A)_j)^n = \sum_{i_1,k_1,...,i_{n},k_{n},i_{n+1}}\left(\prod_{j=1}^{n} c_{i_{j},k_j,i_{j+1},k_j}\right)
\ket{i_{1}}_A\bra{i_{n+1}}_A.
\end{equation}
Taking the trace then yields 
\begin{equation}
\operatorname{Tr}\left(((\rho_A)_j)^n\right) = \sum_{i_1,k_1,...,i_{n},k_{n}}\left(\prod_{j=1}^{n} c_{i_{j},k_j,i_{j+1},k_j}\right)
\end{equation}
with the identification $i_{n + 1} \equiv i_1$. This shows that 
\begin{equation}
-\frac{1}{1-n}\ln\left(\operatorname{Tr}\left((\rho_A)^n\right)\right) = -\frac{1}{1-n}\ln\left(\operatorname{Tr}\left(\left(S_{A,n}\right)\bigotimes_{l=1}^{n}\rho_l \right)\right)
\end{equation}
and provides a way to compute the $n^{\text{th}}$ order Renyi entropies via the SWAP operator $S_{A,n}$.

\section{Renyi shadow norm}\label{app:RenyiShadowNorm}
The $n^{\text{th}}$ order Renyi entropy can be computed as 
\begin{equation}
-\frac{1}{1-n}\log(\operatorname{Tr}(\rho_A^n)) = -\frac{1}{1-n}\log(\operatorname{Tr}(S_{A,n}(\otimes_{s=1}^n\rho_s)))
\end{equation}
where $S_{A,n}$ acts as $S_{A,n}\ket{i_1,i_2,...,i_n}_A\otimes\ket{j_1,...,j_n}_{A^c} = \ket{i_n,i_1,...,i_{n-1}}_A\otimes\ket{j_1,...,j_n}_{A^c}$ on computational basis states, with $A^c$ the complement of the sub-region $A$. Each copy $\rho_s$ is equal to the original density matrix $\rho$. Naturally, we define the unbiased estimator $\operatorname{Tr}\left((S_{A,n})(\otimes_{s=1}^n \rho_{U_s(t),b_s})\right)$ which satisfies the relation $\mathbb{E}_{U_1,b_1,...,U_n,b_n}\operatorname{Tr}\left((S_{A,n})(\otimes_{s=1}^n \rho_{U_s(t),b_s})\right) = \operatorname{Tr}(S_{A,n}(\otimes_{s=1}^n\rho_s))$.
We have that 
\begin{align}
&\text{Var}\left(\operatorname{Tr}\left((S_{A,1,2,...,n})(\otimes_{s=1}^n \rho_{U_s(t),b_s})\right)\right) = \nonumber \\ &\mathbb{E}_{U_1(t),b_1,...,U_n(t),b_n}\left|\operatorname{Tr}\left((S_{A,n})(\otimes_{s=1}^n \rho_{U_s(t),b_s})\right)\right|^2 -   \left|\operatorname{Tr}(S_{A,n}(\otimes_{s=1}^n\rho_s))\right|^2
\end{align}
Since the eigenvalues $\epsilon$ of $S_{A,n}$ are such that $|\epsilon| < 1$, we may neglect the second terms in the r.h.s of the above equation. The first terms may be expanded to yield
\begin{align}\label{eq:RenyiVariance}
D^n\int \left(\prod_{s=1}^ndU_s(t)\bra{0}K_{U_s(t),\varepsilon}(\rho_s)\ket{0}\right)\left|\operatorname{Tr}\left(S_{A,n}\bigotimes_{l=1}^n\mathcal{M}^{-1}\left(U_l(t)^{\dagger}\ket{0}\bra{0}U_l(t)\right)\right)\right|^2
\end{align}
with $D = q^L$ where $L$ is the system size of a single copy $\rho_j$, and we used the locally scrambling property on all the $U_s(t)$ independently to remove the sums over $b_s$.
Now, we may decompose $\mathcal{M}^{-1}(U_s(t)^{\dagger}\ket{0}\bra{0}U_s(t))$ in the Pauli basis as 
$\sum_{\lambda_s}\frac{\bra{0}U_s(t)P_{\lambda_s}U_s(t)^{\dagger}\ket{0}P_{\lambda_s}^{\dagger}}{D\beta_{\lambda_s,\varepsilon}}$. 
Furthermore, we may decompose $S_{A,n}$ as $S_{A,n} = \sum_{i_1,i_2,...,i_n}s_{i_1,i_2,...,i_n}\bigotimes_{s=1}^n P_{i_s}$ for some coefficients $s_{i_1,i_2,...,i_n}$ where the Pauli operator $P_{i_s}$ acts on the $s^{th}$ copy $\rho_s$. Replacing everything in Eq. (\ref{eq:RenyiVariance}) yields

\begin{align}\label{eq:expandedRenyiNorm}
&\frac{1}{D^n}\int \left(\prod_{s=1}^ndU_s(t)\bra{0}K_{U_s(t),\varepsilon}(\rho_s)\ket{0}\right)\times \nonumber \\ 
&\left|\sum_{i_1,...,i_n,\lambda_1,...,\lambda_n}\frac{s_{i_1,...,i_n}}{\prod_{s=1}^n \beta_{\lambda_s,\varepsilon}}\left(\prod_{l=1}^n\bra{0}U_l(t)P_{\lambda_l}U_l(t)^{\dagger}\ket{0} \operatorname{Tr}\left(P_{i_l}P_{\lambda_l}^{\dagger}\right)\right)\right|^2.
\end{align}
Now the only term in an expansion of $K_{U_s(t),\varepsilon}(\rho_s)$ that contributes to the integral is $I/D$, as otherwise, by making use of the locally scrambling property, it can be shown that each term which is not the identity in the decomposition of $K_{U_s(t),\varepsilon}(\rho_s)$ can be made equal to another integral with a different phase provided an appropriately chosen unitary rotation $V$ which is only possible if the result was equal to 0 in the first place. Furthermore, due to the trace, the integral is only non-zero if $\lambda_l = i_l$ for all $l$ from which we obtain that Eq. (\ref{eq:expandedRenyiNorm}) reduces to
\begin{align}
\sum_{i_1,...,i_n,j_1,...,j_n}\frac{s_{i_1,...,i_n}s_{j_1,...,j_n}^*}{\left(\prod_{s=1}^n\beta_{i_s,\varepsilon}\right)\left(\prod_{l=1}^n\beta_{j_l,\varepsilon}\right)}\prod_{r=1}^n \int dU_{r}(t) \bra{0}U_r(t)P_{i_r}U_r(t)^{\dagger}\ket{0}\bra{0}U_r(t)P_{j_r}^{\dagger}U_r(t)^{\dagger}\ket{0}.
\end{align}
Now, by making use of Eq. (\ref{eq:HaarIntegral}), we realize that for the integral not to vanish, it must be the case that $i_r= j_r$ for all $r$, which finally yields 
\begin{align}
\sum_{i_1,...,i_n}\frac{|s_{i_1,...,i_n}|^2}{\left(\prod_{s=1}^n\beta_{i_s,\varepsilon}\right)^2}\prod_{r=1}^n \int dU_{r}(t) |\bra{0}U_r(t)P_{i_r}U_r(t)^{\dagger}\ket{0}|^2.
\end{align}
We know from the calculation of the shadow norm without noise that $\int dU(t) \left|\bra{0}U(t)P_{i}U(t)^{\dagger}\ket{0}\right|^2 = \beta_{i}$ which leads to  
\begin{equation}\label{eq:ExplicitRenyiExpression}
\mathbb{E}_{U_1(t),b_1,...,U_n(t),b_n}\left|\operatorname{Tr}\left((S_{A,n})(\otimes_{s=1}^n \rho_{U_s(t),b_s})\right)\right|^2 = \sum_{i_1,...,i_n}|s_{i_1,...,i_n}|^2\prod_{s=1}^n\frac{\beta_{i_s}}{\left(\beta_{i_{s,\varepsilon}}\right)^2}.
\end{equation}
We note that for any operator $O$ with the Pauli decomposition $O = \sum_{i_1,i_2,...,i_n}s_{i_1,i_2,...,i_n}\bigotimes_{s=1}^n P_{i_s}$, the above result holds. We may now compute the shadow norm at $t = 0$, which yields
\begin{align}\label{eq:t=0Swapeigen}
\sum_{i_1,i_2,...,i_n}\left|s_{i_1,...,i_n}\right|^2\prod_{s=1}^n(q+1)^{\norm{\vec{w}_{i_s}}^2}.
\end{align}
The swap operator $S_{A,n}$ can be written down as a tensor product of SWAP operators over each site of the sub-region $A$, yielding $S_{A,n} = \bigotimes_{i=1}^{|A|} S_{A_i,n}$ where $A_i$ is the $i^{\text{th}}$ site in the sub-region $A$. Let us denote by $g(a,b)$ a function that is equal to $0$ if $(a \text{ mod } q) =  (b \text{ mod } q) = 0$ and is equal to $1$ otherwise. Then by making use of Eq. (\ref{eq:SwapCoefficients}), we find that Eq. (\ref{eq:t=0Swapeigen}) can be written as
\begin{align}
&\prod_{j=1}^{|A|}\sum_{n_{A_j,1},m_{A_j,1},...,n_{A_j,n-1},m_{A_j,n-1}}^{q}\frac{1}{q^{2(n-1)}}(q+1)^{g(n_{A_j,1},m_{A_j,1})}\left(\prod_{s=1}^{n-2}(q+1)^{g(n_{A_j,s+1}-n_{A_j,s},m_{A_j,s+1}-m_{A_j,s})}\right)\times\nonumber \\
&(q+1)^{g(n_{A_j,n-1},m_{A_j,n-1})} \nonumber \\
& = \left(\sum_{n_{1},m_{1},...,n_{n-1},m_{n-1}}^{q}\frac{1}{q^{2(n-1)}}(q+1)^{g(n_{1},m_{1})}\left(\prod_{s=1}^{n-2}(q+1)^{g(n_{s+1}-n_{s},m_{s+1}-m_{s})}\right)(q+1)^{g(n_{n-1},m_{n-1})}\right)^{|A|}.
\end{align} Now, we define the matrix $G$ with matrix elements $G^{n,m}_{k,l} = 1$ if $((k - n) \mod{q}) =  ((l - m) \mod{q}) = 0$, $(q+1)$ otherwise, as well as the vector $V$ with elements $V_{n,m} = 1$  if $(n \mod{q}) = (m \mod{q}) = 0$, $q+1$ otherwise. Using this, we see that the above sum may be written as 
\begin{align}\label{eq:MatrixEigen}
\frac{1}{q^{2(n-1)}}V^T\cdot G^{n-2}\cdot V
\end{align} 
where matrix $G$ takes the form
\begin{align}
G = \underbrace{\left(\begin{array}{cccc}
1 & q+1 & ... & q+1 \\
q+1 & \ddots &  & \\
 \vdots &  &  & q+1 \\
q+1 &  ... & q+1 & 1
\end{array}\right)}_{q^2}
\end{align}
and $V$ takes the form
\begin{align}
V = \left(\begin{array}{c}
1 \\
q+1 \\
\vdots \\
q+1
\end{array}\right).
\end{align}
Now we note that $G = Q - \mathbb{1}q$ where
\begin{align}
Q = \left(\begin{array}{cccc}
q+1 & q+1 & ... & q+1 \\
q+1 & \ddots &  & \\
 \vdots &  &  & q+1 \\
q+1 &  ... & q+1 & q+1
\end{array}\right)
\end{align}
from which we have that 
\begin{align}
G^{n-2} = \sum_{l=0}^{n-2}\binom{n-2}{l}Q^l(-q)^{n-l - 2}.
\end{align}
We further note that $Q^l = q^{2(l-1)}(q+1)^{l-1}Q$ for $l \geq 1$ which yields 
\begin{align}
&\frac{1}{q^{2(n-1)}}V^T\cdot G^{n-2}\cdot V = \nonumber \\ &\frac{1}{q^{2(n-1)}}\sum_{l=0}^{n-2}\binom{n-2}{l}V^{T}QVq^{2(l-1)}(q+1)^{l-1}(-q)^{n-l-2} - \frac{1}{q^{2n-2}}\left(V^{T}QV(q+1)^{-1}q^{-2} - V^{T}V\right)(-q)^{n-2}. 
\end{align}
Now we have that $V^{T}QV = q^2 (q+1) \left(q^2+q-1\right)^2$ and $V^{T}V = 1 + (q^2-1)(q+1)$ from which we finally obtain
\begin{align}\label{eq:matrixProductResult}
\frac{1}{q^{2(n-1)}}V^T\cdot G^{n-2}\cdot V = \frac{\left(q^2(q+1) - q\right)^{n} + (q^2-1)(-q)^n}{q^{2n}}.
\end{align}
The full result is then obtained by raising Eq. (\ref{eq:matrixProductResult}) to the power of $|A|$, which finally yields 
\begin{align}\label{eq:SwapVarT=0}
\left.\mathbb{E}_{U_1(t),b_1,...,U_n(t),b_n}\left|\operatorname{Tr}\left(S_{A,n}(\otimes_{s=1}^n \rho_{U_s(t),b_s})\right)\right|^2 \right|_{t=0} =  \left(\frac{\left(q^2-1\right) (-q)^{n}+\left((q+1) \left(q^2-1\right) + 1\right)^{n}}{q^{2n}}\right)^{|A|}.
\end{align}
To obtain the noise threshold, we must compute the shadow norm at $t = 1$. In order to perform the calculation, we start by counting the number of operators in the Pauli decomposition of $S_{A,n}$ that have local weights $w_{j,s} \in \{0,1\}$ where $j$ is the site index for the $s^{\text{th}}$ copy. We see that the total number of operators associated with the configuration $w_{A_j,s} $, $j \in \{1,2,...,|A|\}$, $s \in \{1,2,...,n\}$ is given by
\begin{align}
&\prod_{j=1}^{|A|}\sum_{n_{A_j,1},m_{A_j,1},...,n_{A_j,n-1},m_{A_j,n-1}}^{q}l(n_{A_j,1},m_{A_j,1})\left(\prod_{s=1}^{n-2}p(n_{A_j,s+1}-n_{A_j,s},m_{A_j,s+1}-m_{A_j,s})\right)\times\nonumber \\
&r(n_{A_j,n-1},m_{A_j,n-1}) 
\end{align}
where $l(n_{A_j,1},m_{A_j,1}) = (1 - w_{A_j,1})$ if $(n_{A_j,s} \text{ mod } q) = (m_{A_j,s} \text{ mod } q) = 0$ and $w_{A_j,1}$ otherwise, $p(n_{A_j,s+1} - n_{A_j,s},m_{A_j,s+1} - m_{A_j,s}) = (1 - w_{A_j,s+1})$ if $((n_{A_j,s+1} - n_{A_j,s}) \text{ mod } q) = ((m_{A_j,s+1} - m_{A_j,s}) \text{ mod } q) = 0$ and $w_{A_j,s+1}$ otherwise, and $r(n_{A_j,n-1},m_{A_j,n-1}) = (1 - w_{A_j,n})$ if $(n_{A_j,n} \text{ mod } q) = (m_{A_j,n} \text{ mod } q) = 0$ and $w_{A_j,n}$ otherwise. This may be written as 
\begin{align}
\prod_{j= 1}^{|A|}L^T\cdot G_{A_j,1}\cdot...\cdot G_{A_j,n-2}\cdot R
\end{align}
with 
\begin{align}
G_{A_j,s} = \underbrace{\left(\begin{array}{cccc}
(1-w_{A_j,s+1}) & w_{A_j,s+1} & ... & w_{A_j,s+1} \\
w_{A_j,s+1} & \ddots &  & \\
 \vdots &  &  & w_{A_j,s+1} \\
w_{A_j,s+1} &  ... & w_{A_j,s+1} & (1-w_{A_j,s+1})
\end{array}\right)}_{q^2}
\end{align}
and $L,R$ take the form
\begin{align}
L = \left(\begin{array}{c}
1 - w_{A_j,1} \\
w_{A_j,1} \\
\vdots \\
w_{A_j,1}  
\end{array}\right), \quad R = \left(\begin{array}{c}
1 - w_{A_j,n} \\
w_{A_j,n} \\
\vdots \\
w_{A_j,n}  
\end{array}\right).
\end{align}
Now we note that the matrices $G_{A_j,s}$ mutually commute and thus the product $G_{A_j,1}\cdot...\cdot G_{A_j,n-2}$ may be computed in any order. Now on the copies $s = 2,...,n-1$ at site $A_j$ there is exactly $\sum_{l=2}^{n-1}w_{A_j,l}$ occupied sites, and there is $n-2 - \sum_{l=2}^{n-1}w_{A_j,l}$ unoccupied sites, from which we find that 
\begin{align}
G_{A_j,1}\cdot...\cdot G_{A_j,n-2} = \left(\begin{array}{cccc}
0 & 1 & ... & 1 \\
1 & \ddots &  & \\
 \vdots &  &  & 1 \\
1 &  ... & 1& 0
\end{array}\right)^{\sum_{l=2}^{n-2}w_{A_j,s}}.
\end{align}
Now by defining the matrix 
\begin{align}
M = \left(\begin{array}{cccc}
1 & 1 & ... & 1 \\
1 & \ddots &  & \\
 \vdots &  &  & 1 \\
1 &  ... & 1& 1
\end{array}\right)
\end{align}
we have that 
\begin{align}
\prod_{j= 1}^{|A|}L^T\cdot G_{A_j,1}\cdot...\cdot G_{A_j,n-2}\cdot R = \prod_{j= 1}^{|A|}L^T\cdot (M - \mathbb{1})^{\sum_{s = 2}^{n-2}w_{A_j,s}}\cdot R.
\end{align}
This is directly analogous to Eq. (\ref{eq:MatrixEigen}) and may thus be computed with the same method. We have that $M^l = q^{2(l-1)}M$ for $l \geq 1$, that $L^{T}MR = ((1-w_{A_j,1}) + w_{A_j,1}(q^2-1))((1-w_{A_j,n}) + w_{A_n,1}(q^2-1)) = (q^2-1)^{w_{A_j,1} + w_{A_j,n}}$ and that $L^{T}R = ((1-w_{A_j,1})(1-w_{A_j,n}) + (q^2 - 1)w_{A_j,1}w_{A_j,n}$
from which we find

\begin{equation}\label{eq:NumberOfPauliInSwap}
\prod_{j= 1}^{|A|}L^T\cdot (M - \mathbb{1})^{\sum_{s = 2}^{n-2}w_{A_j,s}}\cdot R = \prod_{j=1}^{|A|}\left[\frac{(q^2-1)^{\sum_{s=1}^{n}w_{A_j,s}} + (q^2-1)(-1)^{\sum_{s=1}^n w_{A_j,s}}}{q^2}\right].
\end{equation}

\begin{table}
\begin{tabular}{|c|c|c|c|c|c|c|c|c|c|}
\hline
 \text{} & \text{n = 2} & \text{n = 3} & \text{n = 4} & \text{n = 5} & \text{n = 6} & \text{n = 7} & \text{n = 8} & \text{n = 9} & \text{n = 10} \\
 \hline
 \text{ q = 2 } & 0.9396 & 0.9564 & 0.9535 & 0.9539 & 0.9539 & 0.9539 & 0.9539 & 0.9539 & 0.9539 \\
 \text{ q = 3 } & 0.9462 & 0.9551 & 0.9545 & 0.9545 & 0.9545 & 0.9545 & 0.9545 & 0.9545 & 0.9545 \\
 \text{ q = 4 } & 0.9544 & 0.9598 & 0.9596 & 0.9596 & 0.9596 & 0.9596 & 0.9596 & 0.9596 & 0.9596 \\
 \text{ q = 5 } & 0.9609 & 0.9646 & 0.9645 & 0.9645 & 0.9645 & 0.9645 & 0.9645 & 0.9645 & 0.9645 \\
 \text{ q = 6 } & 0.9659 & 0.9686 & 0.9686 & 0.9686 & 0.9686 & 0.9686 & 0.9686 & 0.9686 & 0.9686 \\
 \text{ q = 7 } & 0.9699 & 0.9719 & 0.9719 & 0.9719 & 0.9719 & 0.9719 & 0.9719 & 0.9719 & 0.9719 \\
 \text{ q = 8 } & 0.9731 & 0.9747 & 0.9747 & 0.9747 & 0.9747 & 0.9747 & 0.9747 & 0.9747 & 0.9747 \\
 \text{ q = 9 } & 0.9756 & 0.9769 & 0.9769 & 0.9769 & 0.9769 & 0.9769 & 0.9769 & 0.9769 & 0.9769 \\
 \hline
\end{tabular}
\caption{Values of $f_{\text{th}}$ for the $n^{\text{th}}$ order Renyi entropy as a function of the qudit dimension $q$}
\label{tab:bigfvalues}
\end{table}
Expanding this expression leads to
\begin{equation}
\sum_{s_1,...,s_{|A|} \in \{0,1\}}\frac{(q^2 - 1)^{s_1 + ...+s_{|A|}}}{q^{2|A|}}\prod_{j=1}^{|A|}\left((q^2-1)^{\sum_{s=1}^{n}w_{j,s}}\right)^{1-s_j}\left((-1)^{\sum_{s=1}^{n}w_{j,s}}\right)^{s_j}.
\end{equation}
Now in Eq. (\ref{eq:ExplicitRenyiExpression}) we see that a weight configuration $w_{A_j,s}$ will contribute a product of eigenvalues given by $\prod_{j=1}^A\prod_{s=1}^n\frac{\beta_{(w_{A_j,s},w_{A_2,s},...,w_{A_{|A|},s})}}{(\beta_{\varepsilon, (w_{A_j,s},w_{A_2,s},...,w_{A_{|A|},s})})^2}$. By using our previous calculation of the number of Pauli operators with a weight configuration $w_{A_j,s}$, we find via a sum over all possible weight configurations that
\begin{align}\label{eq:ExactShadowNorm}
&\mathbb{E}_{U_1(t),b_1,...,U_n(t),b_n}\left|\operatorname{Tr}\left(S_{A,n}(\otimes_{s=1}^n \rho_{U_s(t),b_s})\right)\right|^2 = \nonumber \\
&\sum_{s_1,...,s_{|A|} \in \{0,1\}}\frac{(q^2 - 1)^{s_1 + ...+s_{|A|}}}{q^{2n|A|}}\left(\sum_{w_{A_1,1},...,w_{A_{|A|},1}}\left(\prod_{j=1}^{|A|}\left((q^2-1)^{w_{A_j,1}}\right)^{1-s_j}\left((-1)^{w_{A_j,  1}}\right)^{s_j}\right)\frac{\beta_{(w_{A_1,1},w_{A_2,1},...,w_{A_{|A|},1})}}{(\beta_{\varepsilon,(w_{A_1,1},...,w_{A_{|A|},1})})^2} \right)^n.
\end{align}
At $t = 1$, the eigenvalues decompose in pairs as $\beta_{w_{A_1,1},...,w_{A_{|A|,1}}} = \prod_{l=1}^{|A|/2}\beta_{w_{A_{2l-1},1},w_{A_{2l},1}}$. By making use of the fact that at $t = 1$ we have $\beta_{0,0} = 1$, $\beta_{1,0} = \beta_{0,1} = \alpha/f^2$ and $\beta_{1,1} = \alpha/f^4$ with 
\begin{align}\label{eq:ExpandedResult}
\alpha = \frac{2a\left(\frac{1}{q+1}\right) + (1-2a)\left(\frac{1}{q+1}\right)^2}{\left(2a\left(\frac{f}{q+1}\right) + (1-2a)\left(\frac{f}{q+1}\right)^2\right)^2} = \frac{(q+1)^2 \left(q^2+1\right)}{f^2 (f (q-1)+2)^2},
\end{align}
we may perform the sum to find
\begin{align}
&\left.\mathbb{E}_{U_1(t),b_1,...,U_n(t),b_n}\left|\operatorname{Tr}\left(S_{A,1,2,...,n}(\otimes_{s=1}^n \rho_{U_s(t),b_s})\right)\right|^2\right|_{t=1} = \nonumber \\
&\frac{\left(\left(1 + \frac{\alpha}{f^2}(q^2-1)(2 + (q^2-1)\frac{1}{f^2})\right)^n + 2(q^2-1)\left(1 + \frac{\alpha}{f^2}(q^2-2 - (q^2-1)\frac{1}{f^2})\right)^n + (q^2-1)^2\left(1 + \frac{\alpha}{f^2}(\frac{1}{f^2}-2)\right)^n\right)^{|A|/2}}{q^{2n|A|}}.
\end{align}
This is sufficient to compute the thresholds $f_{\text{th}}$ as a function of $q,n$, whose values are displayed in Tab. \ref{tab:bigfvalues}.

\section{Alternate estimators}
To compute the shadow norm of Renyi entropies, we made use of the unbiased estimators $\operatorname{Tr}\left((S_{A,n})(\otimes_{s=1}^n \rho_{U_s(t),b_s})\right)$, but it should be noted that there exist other valid choices. For instance, one may consider the symmetrization \cite{HUANG2020_ClassicalShadow}
\begin{align}\label{eq:SymEstimator}
o_{N} = \frac{1}{N(N-1)...(N-n+1)}\sum_{i_1\neq i_2 ... \neq i_n}^{N}\operatorname{Tr}\left((S_{A,n})(\otimes_{s=1}^n \rho_{U_{i_s}(t),b_{i_s}})\right)
\end{align}
with $N
\geq n$. The first thing to note is that 
\begin{align}
\text{Var}(o_N) \leq \text{Var}\left(\operatorname{Tr}\left((S_{A,n})(\otimes_{s=1}^n \rho_{U_s(t),b_s})\right)\right)
\end{align}
and thus the variance calculated with the estimator $\operatorname{Tr}\left((S_{A,n})(\otimes_{s=1}^n \rho_{U_s(t),b_s})\right)$ in this work is an upper-bound on $\text{Var}(o_N)$. Now we have that for any operator $O$, the estimator $o_{N} = \frac{1}{N(N-1)...(N-n+1)}\sum_{i_1\neq i_2 ... \neq i_n}^{N}\operatorname{Tr}\left(O(\otimes_{s=1}^n \rho_{U_{i_s}(t),b_{i_s}})\right)$ is invariant under permutations of the $\rho_{U_{i_s},b_{i_s}}$ such that we may replace $O$ by $O_s = \frac{1}{n!}\sum_{\sigma_s \in \mathcal{S}_n}\sigma_s(O)$ where $\sigma_s$ acts on a tensor product of Pauli operators as $\sigma_s(P_1\otimes P_2 \otimes ... \otimes P_n) = P_{\sigma_s(1)} \otimes P_{\sigma_s(2)} ... \otimes P_{\sigma_s(n)}$ and $\mathcal{S}_n$ is the set of all permutations of $n$ elements, from which we may expand $\text{Var}(o_N)$ to obtain
\begin{align}\label{eq:SymmetrizedVariacne}
&\text{Var}(o_N) = \nonumber \\ &\binom{N}{n}^{-2}\sum_{l=1}^{n}\left(\binom{N}{n}\binom{n}{l}\binom{N-n}{n-l}\mathbb{E}_{U_1,b_1,...,U_{l},b_{l}}\left|\operatorname{Tr}\left(O_s(\otimes_{s=1}^{l} \rho_{U_{s}(t),b_{s}})\otimes_{j=1}^{n-l}\rho\right)\right|^2 - \right. 
\left.\left|\operatorname{Tr}\left(O_s\otimes_{j=1}^n\rho\right)\right|^2\right) \nonumber \\
&= \binom{N}{n}^{-2}\sum_{l=1}^{n}\binom{N}{n}\binom{n}{l}\binom{N-n}{n-l}\text{Var}\left(\operatorname{Tr}\left(O_s(\otimes_{s=1}^{l} \rho_{U_{s}(t),b_{s}})\otimes_{j=1}^{n-l}\rho\right)\right)  \nonumber \\
\end{align}
where we have used the fact that $O_s$ is symmetric under arbitrary permutations to recast the estimators in the ordered tensor product $\otimes_{s=1}^{l} \rho_{U_{i_s}(t),b_{i_s}}\otimes_{j=1}^{n-l}\rho$ 
without changing the value of the expectation value. Let us now focus on the calculation of the Renyi entropies where the operator $O = S_{A,n}$. In this case the terms proportional to $\left|\operatorname{Tr}\left(S_{A,n}\otimes_{j=1}^n\rho\right)\right|^2 \leq 1$ are of $O(1)$ and can thus be ignored. To obtain $\text{Var}(o_N)$, we must evaluate $\mathbb{E}_{U_1,b_1,...,U_{l},b_{l}}\left|\operatorname{Tr}\left(O_s(\otimes_{s=1}^{l} \rho_{U_{s}(t),b_{s}})\otimes_{j=1}^{n-l}\rho\right)\right|^2$. Following the calculations of the previous sections, we find that 
\begin{align}
&\mathbb{E}_{U_1,b_1,...,U_{l},b_{l}}\left|\operatorname{Tr}\left(O_s(\otimes_{s=1}^{l} \rho_{U_{s}(t),b_{s}})\otimes_{j=1}^{n-l}\rho\right)\right|^2 = \frac{1}{D^l}\int \left(\prod_{s=1}^l dU_s(t)\bra{0}K_{U_s(t),\varepsilon}(\rho_s)\ket{0}\right)\times \nonumber \\ 
&\left|\sum_{i_1,...,i_n,\lambda_1,...,\lambda_l}\frac{\alpha_{i_1,...,i_n}}{\prod_{s=1}^l \beta_{\lambda_s,\varepsilon}}\left(\prod_{r=1}^l\bra{0}U_r(t)P_{\lambda_r}U_r(t)^{\dagger}\ket{0} \operatorname{Tr}\left(P_{i_r}P_{\lambda_r}^{\dagger}\right)\right)\prod_{t=l+1}^{n-l}\operatorname{Tr}\left(P_{i_t}\rho\right)\right|^2,
\end{align}

where the $\alpha_{i_1,...,i_n}$ are the Pauli coefficients of the symmetrized $S_{A,n}$ operator $\left(\frac{1}{n!}\sum_{\sigma_s \in S_n}\sigma_s(S_{A,n})\right)$. Expanding, we find

\begin{align}
\sum_{i_1,...,i_n,j_1,...,j_n}\frac{\alpha_{i_1,...,i_n}\alpha_{j_1,...,j_n}^*}{\left(\prod_{s=1}^l\beta_{i_s,\varepsilon}\right)\left(\prod_{s=1}^l\beta_{j_s,\varepsilon}\right)}\left(\prod_{r=1}^l \int dU_{r}(t) \bra{0}U_r(t)P_{i_r}U_r(t)^{\dagger}\ket{0}\bra{0}U_r(t)P_{j_r}^{\dagger}U_r(t)^{\dagger}\ket{0}\right) \times \nonumber\\
\left(\prod_{t=l+1}^{n-l}\operatorname{Tr}\left(P_{i_t}\rho\right)\operatorname{Tr}\left(P_{j_t}^{\dagger}\rho\right)\right),
\end{align}
which reduces to 
\begin{align}
\sum_{i_{l+1},...,i_n,j_{l+1},...,j_n}\left(\sum_{i_1,...,i_l} \alpha_{i_1,...,i_n}\alpha_{i_1,...,i_l,j_{l+1},...,j_n}^*\prod_{s=1}^l\frac{\beta_{i_s}}{\left(\beta_{i_{s,\varepsilon}}\right)^2}\right)\left(\prod_{t=l+1}^{n}\operatorname{Tr}\left(P_{i_t}\rho\right)\operatorname{Tr}\left(P_{j_t}^{\dagger}\rho\right)\right).
\end{align}
To make further progress, we note that $|\alpha_{i_1,...,i_n}\alpha_{j_1,...,j_n}^*| \leq \frac{1}{q^{2|A|(n-1)}}$ and that the coefficients are only non-zero for Pauli operators that have support within $A$, as other Pauli operators do not appear in the decomposition of $S_{A,n}$. Thus we obtain the upper-bound
\begin{align}
&\mathbb{E}_{U_1,b_1,...,U_{l},b_{l}}\left|\operatorname{Tr}\left(O_s(\otimes_{s=1}^{l} \rho_{U_{s}(t),b_{s}})\otimes_{j=1}^{n-l}\rho\right)\right|^2 \leq \nonumber \\
&\left( \frac{1}{q^{2|A|(n-1)}}\left(\sum_{i:\text{Sup}(P_i) \in A}\frac{\beta_{i}}{\left(\beta_{i,\varepsilon}\right)^2}\right)^l\right)\left|\sum_{i:\text{Sup}(P_i) \in A}\operatorname{Tr}(P_{i}\rho)\right|^{2(n-l)} \leq 
\nonumber \\
&\left(\frac{\sum_{i:\text{Sup}(P_i) \in A}\frac{\beta_{i}}{\left(\beta_{i,\varepsilon}\right)^2}}{q^{2|A|}}\right)^lq^{2|A|}
\end{align}
where we used the fact that $|\sum_{i:\text{Sup}(P_i) \in A}\operatorname{Tr}(\rho P_i)| = |\operatorname{Tr}(\rho \left(\sum_{i:\text{Sup}(P_i) \in A} P_i\right))| \leq q^{|A|}$ since the operator $\left(\sum_{i:\text{Sup}(P_i) \in A} P_i\right)$ given in Eq. (\ref{eq:maxEigenSum}) has a maximum eigenvalue of $q^{|A|}$ and $\text{Sup}(P_i) $ is the support of the Pauli operator $P_i$. The sum $\left(\frac{\sum_{i:\text{Sup}(P_i) \in A}\frac{\beta_{i}}{\left(\beta_{i,\varepsilon}\right)^2}}{q^{2|A|}}\right)$ may be recast as $\sim q^{\alpha|A|}$ for some positive $\alpha$. This suggests that the case $l = n$ dominates at large $|A|$, which corresponds to the estimator used in this work, and may thus be used as a good approximation for the threshold values $f_{\text{th}}$ obtained with a symmetric estimator $o_N$ in the limit of large $|A|$. For $t= 0,1$ we find explicitly
\begin{align}
\left.\left(\sum_{i:\text{Sup}(P_i) \in A}\frac{\beta_{i}}{\left(\beta_{i,\varepsilon}\right)^2}\right)^l\right|_{t = 0} = ((q^2-1)(q+1) + 1)^{|A|l}
\end{align}
and
\begin{align}
\left.\left(\sum_{i:\text{Sup}(P_i) \in A}\frac{\beta_{i}}{\left(\beta_{i,\varepsilon}\right)^2}\right)^l\right|_{t = 1} = \left(1 + 2\frac{\alpha}{f^2}(q^2 - 1) + (q^2 - 1)^2\frac{\alpha}{f^4}\right)^{\frac{|A|}{2}l} 
\end{align}
from which we deduce
\begin{align}
&\left.\mathbb{E}_{U_1,b_1,...,U_{l},b_{l}}\left|\operatorname{Tr}\left(O_s(\otimes_{s=1}^{l} \rho_{U_{s}(t),b_{s}})\otimes_{j=1}^{n-l}\rho\right)\right|^2\right|_{t=0} \leq \nonumber \left(\frac{((q^2-1)(q+1) + 1)}{q^2}\right)^{|A|l}q^{2|A|}
\end{align}
\begin{align}
&\left.\mathbb{E}_{U_1,b_1,...,U_{l},b_{l}}\left|\operatorname{Tr}\left(O_s(\otimes_{s=1}^{l} \rho_{U_{s}(t),b_{s}})\otimes_{j=1}^{n-l}\rho\right)\right|^2\right|_{t=1} \leq \nonumber \left(\frac{\left(1 + 2\frac{\alpha}{f^2}(q^2 - 1) + (q^2 - 1)^2\frac{\alpha}{f^4}\right)^{1/2}}{q^2}\right)^{|A|l}q^{2|A|}.
\end{align}
Replacing in Eq. (\ref{eq:SymmetrizedVariacne}) then yields an upper-bound on $\text{Var}(o_N)$ at $t = 0,1$. We further note that the absolute thresholds such as $t_{\text{max}}$ and the lower bound on $f_{\text{th}}$ remain unaffected as even in the symmetrisation scheme, the overall variance is given by an appropriate sum of products of the shadow norm of individual Pauli operators, and the absolute thresholds apply to the shadow norm of any Pauli operator.

\section{Alternate post-processing scheme}
It is natural to ask whether it is possible to improve or mitigate the noise in the post-processing step.
Let us assume that we apply single-qubit trace-preserving Pauli maps $\kappa$ between each unitary layer of the post-processing stage in an attempt to mitigate the effect of noise. We represent the application of $\kappa$ on all qubits of the system as $\kappa_L = \otimes^L \kappa$. We further assume that the noise in the system $\varepsilon$ is a uniform diagonal map of the Pauli operators, i.e $\varepsilon(X^nZ^m) = f_{\varepsilon}X^nZ^m$. In this setting, the channel $\mathcal{M}$ becomes
\begin{align}
\mathcal{M}(\rho)
= D\int_{\mathcal{U}(t)} dU(t)\bra{0}K_{\varepsilon,U(t)}(\rho) \ket{0} K_{\kappa,U^{\dagger}(t)}\left(\ket{0}\bra{0}\right),
\end{align}
where
\begin{align}
K_{\kappa,U^{\dagger}(t)}\left(\ket{0}\bra{0}\right) = U_0^{\dagger}k_L(U_1^{\dagger}...k_L(U_0'^{\dagger} \ket{0}\bra{0}U_0' )... U_1)U_0.
\end{align}
We start by calculating the eigenvalue
\begin{align}
\beta_{i,\varepsilon, \kappa} = \frac{1}{D}\operatorname{Tr}(P_{i}^{\dagger}\mathcal{M}(P_{i})).
\end{align} 
Assuming the Krauss decomposition $\kappa(.) = \sum_j K_j (.) J_j^{\dagger}$, we define $\kappa^{-1}(.) \equiv \sum_j J_j^{\dagger} (.) K_j$. With this, we find that 
\begin{align}
\beta_{i,\varepsilon, \kappa} =  \int_{\mathcal{U}(t)} dU(t)\bra{0}K_{\varepsilon,U(t)}(P_{i}) \ket{0}\bra{0} K_{\kappa^{-1},U(t)}(P_{i}^{\dagger})\ket{0}.
\end{align} 
The choice of $\kappa^{-1}$ is arbitrary, but the assumption that $\varepsilon$ is a diagonal Pauli map implies that only its diagonal elements, which we denote by $f_{(n,m),\kappa^{-1}}$, will contribute. Now, we may decompose $K_{\varepsilon,U(t)}(P_{i})$ in the Pauli basis to get $K_{\varepsilon,U(t)}(P_{i}) = \sum_\lambda \alpha_{i,\lambda,U(t),\varepsilon}P_{\lambda}$ and 
$K_{\kappa^{-1}, U(t)}(P_{i}^{\dagger}) = \sum_\lambda \alpha_{i,\lambda,U(t), \kappa^{-1}}^*P_{\lambda}^{\dagger}$. Following closely the calculations for the noiseless case, one obtains 
\begin{align}
\beta_{i,\varepsilon,\kappa} = \sum_{\vec{w}}\left(\sum_{\lambda: \vec{w}_\lambda = \vec{w}} \overline{\alpha^*_{i,\lambda,U(t),\kappa^{-1}}\alpha_{i,\lambda,U(t),\varepsilon}} \right)\frac{1}{(q + 1)^{\norm{\vec{w}}^2}}
\end{align}
where 
\begin{align}
\overline{\alpha^*_{i,\lambda,U(t), \kappa^{-1}}\alpha_{i,\lambda,U(t),\varepsilon}}  = \int dU(t)\alpha^*_{i,\lambda,U(t), \kappa^{-1}}\alpha_{i,\lambda,U(t),\varepsilon}.
\end{align}
Taking advantage of the fact that the multi-qudits Clifford group is a 2-design, we compute the average  with Clifford unitaries

\begin{align}
\beta_{i,\varepsilon, \kappa} =  \sum_{w}\left(\sum_{\lambda: \norm{\vec{w}_\lambda}^2 = w} \frac{1}{|\mathcal{C}(t)|}\sum_{U(t) \in \mathcal{C}(t)}\alpha_{i,\lambda,U(t),\kappa^{-1}}^*\alpha_{i,\lambda,U(t),\varepsilon} \right)\frac{1}{(q + 1)^{w}}
\end{align}
where $\mathcal{C}(t)$ is the set of all circuits of depth $t$ where the local unitaries acting on $n$ qudits of dimension $q$ are selected at random from $\mathcal{C}_n^q$, the Clifford group on $n$ qudits.  We have denoted above the total number of elements in $\mathcal{C}(t)$ by $|\mathcal{C}(t)|$. From here, we see that the calculation is equivalent to the one in the absence of the channel $\kappa$, but with a an effective noise parameter now equal to 
\begin{align}
f = f_{\varepsilon}\sum_{(n,m) \neq (0,0)}\frac{f_{(n,m),\kappa^{-1}}}{q^2-1}.
\end{align}
To get the full picture, we must compute the shadow norm, which is given in this case by 
\begin{align}
\mathbb{E}_{U(t), b}\left[|o_{U(t),b}|^2\right]=  D\int_{\mathcal{U}(t)} \mathrm{d} U(t) \bra{0}K_{\varepsilon,U(t)}(\rho)\ket{0}|\operatorname{Tr}[P_{i} \mathcal{M}^{-1}(K_{\kappa,U^{\dagger}(t)}(\ket{0}\bra{0}) )]|^2.
\end{align}
Expanding and following previous calculations of the shadow norm, one finds 
\begin{align}
\mathbb{E}_{U(t), b}\left[|o_{U(t),b}|^2\right]=  \frac{1}{\beta_{i,\varepsilon,\kappa}^2}\int_{\mathcal{U}(t)} \mathrm{d} U(t) |\bra{0}K_{\kappa^{-1},U(t)}(P_i)\ket{0}|^2 = \frac{\beta_{i,(\kappa^{-1})^2}}{\beta_{i,\varepsilon,\kappa}^2},
\end{align}
where $(\kappa^{-1})^2$ is defined to be the channel $(\kappa^{-1})^2:X^nZ^m \rightarrow (f_{(n,m),\kappa^{-1}})^{2}X^nZ^m$. Treating the coefficients $f_{(n,m),\kappa^{-1}}$ as parameters, one can minimize the variance. To do so, we make use of the relations
\begin{align}
\beta_{i,\varepsilon,\kappa} = 
\sum_{\vec{w}(t)}\left(\sum_{\vec{w}(1),...,\vec{w}(t-1)}\text{Pr}(\vec{w}(0),...,\vec{w}(t))\left(\sum_{(n,m)\neq (0,0)}\frac{f_{\varepsilon}f_{(n,m),\kappa^{-1}}}{(q^2 - 1)}\right)^{w_{\text{tot}} - 1}\right)\frac{1}{(q+1)^{\norm{\vec{w}(t)}^2}}
\end{align}
\begin{align}
\beta_{i,(\kappa^{-1})^2} = 
\sum_{\vec{w}(t)}\left(\sum_{\vec{w}(1),...,\vec{w}(t-1)}\text{Pr}(\vec{w}(0),...,\vec{w}(t))\left(\sum_{(n,m)\neq (0,0)}\frac{f_{(n,m),\kappa^{-1}}^2}{(q^2 - 1)}\right)^{w_{\text{tot}}}\right)\frac{1}{(q+1)^{\norm{\vec{w}(t)}^2}}.
\end{align}
The minimisation condition is given by 
$\frac{\partial}{\partial f_{(l,k),\kappa^{-1}}}\frac{\beta_{i,(\kappa^{-1})^2}}{\beta_{i,\varepsilon,\kappa}^2} = 0$ for all $(l,k)$, which after simplification yields
\begin{align}
\sum_{\vec{w}(t)}\left(\sum_{\vec{w}(1),...,\vec{w}(t-1)}\text{Pr}(\vec{w}(0),...,\vec{w}(t))\left(\sum_{(n,m)\neq (0,0)}\frac{f_{(n,m),\kappa^{-1}}^2}{(q^2 - 1)}\right)^{w_{\text{tot}}}\right)\frac{1}{(q+1)^{\norm{\vec{w}(t)}^2}}\times \nonumber \\ 
\sum_{\vec{w}(t)}\left(\sum_{\vec{w}(1),...,\vec{w}(t-1)}\text{Pr}(\vec{w}(0),...,\vec{w}(t))w_{\text{tot}}\left(\sum_{(n,m)\neq (0,0)}\frac{f_{\varepsilon}f_{(n,m),\kappa^{-1}}}{(q^2 - 1)}\right)^{w_{\text{tot}} - 1}\frac{2f_{\varepsilon}}{q^2-1}\right)\frac{1}{(q+1)^{\norm{\vec{w}(t)}^2}} = \nonumber \\
\sum_{\vec{w}(t)}\left(\sum_{\vec{w}(1),...,\vec{w}(t-1)}\text{Pr}(\vec{w}(0),...,\vec{w}(t))w_{\text{tot}}\left(\sum_{(n,m)\neq (0,0)}\frac{f_{(n,m),\kappa^{-1}}^2}{(q^2 - 1)}\right)^{w_{\text{tot}}-1}\frac{2f_{(l,k),\kappa^{-1}}}{q^2-1}\right)\frac{1}{(q+1)^{\norm{\vec{w}(t)}^2}}\times \nonumber \\ 
\sum_{\vec{w}(t)}\left(\sum_{\vec{w}(1),...,\vec{w}(t-1)}\text{Pr}(\vec{w}(0),...,\vec{w}(t))\left(\sum_{(n,m)\neq (0,0)}\frac{f_{\varepsilon}f_{(n,m),\kappa^{-1}}}{(q^2 - 1)}\right)^{w_{\text{tot}}}\right)\frac{1}{(q+1)^{\norm{\vec{w}(t)}^2}}.
\end{align}
It is easy to see that these conditions are satisfied for all $(l,k)$ provided $f_{(l,k),\kappa^{-1}} = f_{\varepsilon}$, which agrees with the intuition that the inverse of the channel $\mathcal{M}$ should undo the effect of $K_{\kappa^{-1},U(t)}(P_i)$ to mitigate noise maximally. Through simple mean-field calculations discussed in the text, we know that the effective $f$ parameter contributes a factor $\approx f^{\overline{w_{\text{tot}}}}$ to the eigenvalue. Thus, we may compare the new protocol with the original one via the approximate ratio 
\begin{align}
\left(\frac{\beta_{i,(\kappa^{-1})^2} }{\beta_{i,\varepsilon,\kappa}^2 }\right) /\left(\frac{\beta_i}{\beta_{i,\epsilon}^2}\right)  \approx \left(\frac{\left(\sum_{(n,m) \neq (0,0)}\frac{f_{(n,m),\kappa^{-1}}^2}{q^2-1}\right)\left(\sum_{(n,m) \neq (0,0)}\frac{f_{\varepsilon}}{q^2-1}\right)^2}{\left(\sum_{(n,m) \neq (0,0)}\frac{f_{\varepsilon}f_{(n,m),\kappa^{-1}}}{q^2-1}\right)^2}\right)^{\overline{w_{\text{tot}}}}.
\end{align}
Taking $f_{(n,m),\kappa^{-1}} = f_{\varepsilon}$ yields 

\begin{align}
\left(\frac{\beta_{i,(\kappa^{-1})^2} }{\beta_{\varepsilon,\kappa}^2 }\right) /\left(\frac{\beta_i}{\beta_{i,\epsilon}^2}\right) \approx \left(\frac{\left(\sum_{i}\frac{f_{\varepsilon}}{q^2-1}\right)^2}{\left(\sum_{i}\frac{f_{\varepsilon}^2}{q^2-1}\right)}\right)^{\overline{w_{\text{tot}}}}  = 1.
\end{align} 
This suggests that the effect of noise in the case of a uniform depolarizing channel $\varepsilon$ cannot be mitigated, or even improved, via classical post-processing. We postpone the study of more general noise models $\varepsilon$ to future work.


\section{Approximate $t_*$ scaling}\label{app:OptimalDepthAverage}
The approximate bulk density $n(t)$ for an initial Pauli operator with contiguously occupied sites takes the form \cite{IPPOLITI2023_OperatorSpreading} $n(t) = (1-q^{-2}) + g(t)$ where $g(t)$ is some function which in the large $t$ limit behaves as $ct^{-3/2}e^{-\gamma t} + ...$ where the $...$ denote sub-leading terms, $c = \frac{1}{\sqrt{\pi}\gamma(q^2+1)}$  and $\gamma = 2\log(\frac{q^2+1}{2q})$ \cite{IPPOLITI2023_OperatorSpreading}. To find how the optimal depth $t_*$ scales with $k$ for an initial Pauli operator with $k$ contiguous occupied sites, we study an approximate version of the dynamics where we assume that the Pauli operator at time $t$ under the unitary time evolution has size $k + 2v_Bt$ where $v_B = 1-\frac{2}{q^2+1}$ is the butterfly velocity of the dynamics and we assume that each site within the Pauli operator boundaries contains a non-identity Pauli operator at site $j$ with probability $n(t)$ \cite{IPPOLITI2023_OperatorSpreading}. We recall that we may write $\beta_{i,\varepsilon}$ as 
\begin{align}
&\beta_{i,\varepsilon} = 
\sum_{\vec{w}(t)} \left(\frac{q+1}{f}\right)^{-\norm{\vec{w}(t)}^2}
\sum_{\vec{w}(1),...,\vec{w}(t-1)}\text{Pr}(\vec{w}(0),...,\vec{w}(t))\left(\frac{1}{f}\right)^{-\sum_{s=0}^{t-1}\norm{\vec{w}(s)}^2}.
\end{align}
By making use of the above simplification scheme, the probability distribution becomes uncorrelated, i.e $\text{Pr}(\vec{w}(0),...,\vec{w}(t)) = \left(\prod_{j=0}^{t-1}\text{Pr}(\vec{w}(j))\right)\text{Pr}(\vec{w}(t))$. We thus get 
\begin{align}
\beta_{i,\varepsilon} \approx \sum_{\vec{w}(t)}\text{Pr}(\vec{w}(t))\left(\frac{f}{q+1}\right)^{\norm{\vec{w}(t)}^2}\prod_{j=0}^{t-1}\left(\sum_{\vec{w}(j)}\text{Pr}(\vec{w}(j))f^{\norm{\vec{w}(j)}^2}\right).
\end{align}
We have that 
\begin{align}
\sum_{\vec{w}(j)}\text{Pr}(\vec{w}(j))f^{\norm{\vec{w}(j)}^2} = \sum_{w = 0}^{k + 2v_Bj}\binom{k + 2v_Bj}{w}n(j)^{w}(1-n(j))^{k + 2v_Bj - w}f^w = \left(n(j)f + (1-n(j))\right)^{k+2v_Bj}
\end{align}
and 
\begin{align}
\sum_{\vec{w}(t)}\text{Pr}(\vec{w}(t))\left(\frac{f}{q+1}\right)^{\norm{\vec{w}(t)}^2} = \left(n(j)\frac{f}{q+1} + (1-n(j))\right)^{k+2v_Bj}.
\end{align}
Using these results, we readily find that 
\begin{align}
\beta_{i,\varepsilon} \approx \left(n(t)\frac{f}{q+1} + (1-n(t))\right)^{k+2v_Bt}\prod_{j=0}^{t-1}\left(n(j)f + (1-n(j))\right)^{k+2v_Bj}.
\end{align}
The noiseless eigenvalue $\beta_i$ is obtained by taking the limit $f \rightarrow 1$ which leads to 
\begin{align}
\norm{P_i}_{\text{sh}}^2 \approx  \left(\frac{n(t)}{q+1} + (1-n(t))\right)^{(k+2v_Bt)}\left(\frac{n(t)f}{q+1} + (1-n(t))\right)^{-2(k+2v_Bt)}\prod_{j=0}^{t-1}\left(n(j)f + (1-n(j))\right)^{-2(k+2v_Bj)}.
\end{align}
Now we desire to estimate the point at which this expression reaches a minimum. To do so, we start by taking the logarithm of the above, which yields 
\begin{align}
&\ln(\norm{P_i}_{\text{sh}}^2) = -2\sum_{j=0}^{t-1}(k + 2v_Bj)\ln{\left(n(j)f + (1-n(j))\right)} -2(k + 2v_Bt)\ln{\left(\frac{n(t)f}{q+1} + (1-n(t))\right)} +\nonumber \\ 
&(k + 2v_Bt)\ln{\left(\frac{n(t)}{q+1} + (1-n(t))\right)}.
\end{align}
To find the minimum, we compute $\ln(\norm{P_i}_{\text{sh}}^2)|_{t+1} - \ln(\norm{P_i}_{\text{sh}}^2)|_{t}$ and then find the value of $t$ for which this vanishes. Further replacing $n(t)$ by $(1 - q^{-2}) + g(t)$, we find
\begin{align}
-2(k+2v_Bt)\ln\left(\frac{f(q^2-1)+1}{q^2}\left(1 + \frac{g(t)(f-1)q^2}{f(q^2-1)+1}\right)\right) \nonumber \\ 
-2(k + 2v_B(t+1))\ln{\left(\frac{f(q-1)+1}{q^2}\left(1+\frac{g(t+1)(f - (q+1))q^2}{(q+1)(f(q-1)+1)}\right)\right)} \nonumber \\ +2(k + 2v_Bt)\ln{\left(\frac{f(q-1)+1}{q^2}\left(1+\frac{g(t)(f - (q+1))q^2}{(q+1)(f(q-1)+1)}\right)\right)} \nonumber \\ 
+(k + 2v_B(t+1))\ln{\left(\frac{1}{q}\left(1-\frac{g(t+1)q^2}{(q+1)}\right)\right)} -(k + 2v_Bt)\ln{\left(\frac{1}{q}\left(1-\frac{g(t)q^2}{(q+1)}\right)\right)}. 
\end{align}
At large $t$, we have that $g(t) \ll 1$ which we use to expand $\ln(1 + x) \approx x$. Furthermore, at large $t$ one has that $g(t+1) \approx e^{-\gamma} g(t)$. With these approximations we find
\begin{align}
(k + 2v_Bt)g(t)\left(-2\left(\frac{(f-1)q^2}{f(q^2-1)+1}\right)+\left(\left(\frac{q^2}{(q+1)}\right) +2 \left(\frac{(f - (q+1))q^2}{(q+1)(f(q-1)+1)}\right)\right)(1 - e^{-\gamma} )\right) \nonumber\\ =
2(k + 2v_Bt)\left(\ln\left(\frac{f(q^2-1)+1}{q^2}\right)\right) + 2v_B\left(2\ln{\left(\frac{f(q-1)+1}{q^2}\right)} - \ln{\left(\frac{1}{q}\right)}\right)  + O(g(t)).
\end{align}
We are interested in the limit of large $k$. Dividing both sides by $k + 2v_Bt$, we find 
\begin{align}\label{eq:meanfieldtrans}
g(t)\left(-2\left(\frac{(f-1)q^2}{f(q^2-1)+1}\right)+\left(\left(\frac{q^2}{(q+1)}\right) +2 \left(\frac{(f - (q+1))q^2}{(q+1)(f(q-1)+1)}\right)\right)(1 - e^{-\gamma} )\right) \nonumber \\  = 2\left(\ln\left(\frac{f(q^2-1)+1}{q^2}\right)\right) + O(1/k) + O(g(t)/k).
\end{align}
We note that in the noiseless limit, the first term in the r.h.s vanishes and the dominant contribution arises from the $O(1/k)$ term, leading to the noiseless $t_* \approx \ln(k)$ behavior. By contrast, for any finite noise, the first term of the r.h.s always dominates at large enough $k$. By dropping the terms $2v_Bt$ in front of $k$ and ignoring terms that go as $O(g(t)/k)$, the above reduces to an equation of the form 
\begin{align}
e^{-\gamma t}t^{-3/2} = A + \frac{B}{k}.
\end{align}
with 
\begin{align}
A = \frac{2\left(\ln\left(\frac{f(q^2-1)+1}{q^2}\right)\right)}{\left(-2\left(\frac{(f-1)q^2}{f(q^2-1)+1}\right)+\left(\left(\frac{q^2}{(q+1)}\right) +2 \left(\frac{(f - (q+1))q^2}{(q+1)(f(q-1)+1)}\right)\right)(1 - e^{-\gamma} )\right)c}\nonumber \\
B = \frac{2v_B\left(2\ln{\left(\frac{f(q-1)+1}{q^2}\right)} - \ln{\left(\frac{1}{q}\right)}\right)}{\left(-2\left(\frac{(f-1)q^2}{f(q^2-1)+1}\right)+\left(\left(\frac{q^2}{(q+1)}\right) +2 \left(\frac{(f - (q+1))q^2}{(q+1)(f(q-1)+1)}\right)\right)(1 - e^{-\gamma} )\right)c}.
\end{align}
An explicit solution for this equation is given by 
\begin{align}
t_* = \frac{3}{2\gamma}W\left(\frac{2\gamma}{3}\left(\frac{1}{A + B/k}\right)^{2/3}\right),
\end{align}
where $W$ is the Lambert $W$ function, which may be approximated as 
\begin{align}
t_* = \frac{1}{\gamma}\left(\ln\left(\frac{1}{A + B/k}\right)\right)+ O\left(\ln\left(\ln\left(\frac{1}{A + B/k}\right)\right)\right).
\end{align}
We note that in the noiseless limit, $A = 0$ recovers the noiseless $t_* \approx \ln(k)$. For any finite noise, $A$ is non-zero, which leads to the asymptotic behavior $t_* \rightarrow \frac{1}{\gamma}\ln{\frac{1}{A}} + O(\ln(\ln(\frac{1}{A})))$ in the limit $k \rightarrow \infty$. 
\section{Absolute upper bound $t_{\text{max}}$ on $t_*$}
As discussed in the main text, there are three mechanisms contributing to the shadow norm. We have i) bulk decay which decreases the shadow norm, ii) operator growth which increases the shadow norm and iii) noise which increases the shadow norm. By removing the effect of operator growth and by maximizing the effect of bulk decay, we can find the maximum possible value $t_\text{max}$ for $t_*$. This can be achieved by considering a fully occupied Pauli operator with $k = L$ occupied sites in a periodic system of size $L\gg t_*$. 
In this setting there is no operator growth, for times $t\ll L$ this is equivalent to taking the limit $k \rightarrow \infty$ and the effect of bulk decay is maximized, which means that $t_*$ corresponds to $t_{\text{max}}$ in this context. The numerical value of $t_{\text{max}}$ can be computed with the methods discussed in the next section, whose results are displayed in the main text. Furthermore, the shadow norm for an arbitrary operator $O$ is given by an appropriate sum of products of the shadow norm of individual Pauli operators, from which it directly follows that $t_{\text{max}}$ holds for arbitrary operators $O$.

\section{MPS simulations}\label{app:NumericalSimulation}
We note that $\beta_{i,\varepsilon}$ does not depend on which type of Pauli operator is at site $j$ in $P_i$, only the weight of the Pauli operator plays a role in the calculation. The unitary layers composed of single-site unitary operators do not change the weight distribution of $P_i$, and thus act as the identity on the weight configurations $\vec{w}(j)$. The subsequent layers do however change the weight of the Pauli operator $P_i$ and must be accounted for. By making use of the fact that the multi-qudit Clifford group is a two-design, we can perform the average with Clifford unitary gates. A randomly chosen Clifford gate acting on a pair of sites $j,j+1$ sends any non-identity Pauli operator with support on the sites $j,j+1$ to any other non-identity Pauli operator on the sites $j,j+1$ with equal frequency. Out of those, $(q^2 - 1)$ are the identity on the site $j+1$ while hosting some Pauli operator on the site $j$. Similarly, $(q^2 - 1)$ Pauli operators are the identity on the site $j$ and have a Pauli operator on the site $j+1$. Finally, $(q^2 - 1)^2$ of them host non-identity Pauli operators on both sites. Using the fact that each Pauli operator is equally probable, one finds that the first two cases occur with probability $\frac{q^2 - 1}{q^4 - 1}$, whilst the last case occurs with probability $\frac{(q^2 - 1)^2}{q^4 - 1}$. By making use of the symbol $\circ_j$ to denote a site $j$ hosting the identity operator and the symbol $\bullet_j$ to denote a site $j$ hosting a Pauli operator, we have that the above can be summarized as follows; upon averaging over Clifford unitaries acting on sites $j, j+1$, any non-identity Pauli operator acting on the sites $j, j+1$ will have a weight configuration $(\circ_j\bullet_{j+1})$ with probability $\frac{q^2 - 1}{q^4 - 1}$, $(\bullet_j\circ_{j+1})$ with probability $\frac{q^2 - 1}{q^4 - 1}$ and $(\bullet_j\bullet_{j+1})$  with probability $\frac{(q^2 - 1)^2}{q^4 - 1}$. The operator $(\circ_j \circ_{j+1})$ is mapped to itself by the Clifford unitaries. Given a vector $(p_{\circ_j,\circ_{j+1}}^{t},
p_{\circ_j,\bullet_{j+1}}^t,
p_{\bullet_j,\circ_{j+1}}^t,
p_{\bullet_j,\bullet_{j+1}}^t)$ containing the probabilities to find either of the possible weight configurations at layer $t$, one may describe the action of the two-sites Clifford unitary using an update matrix to find the probabilities at layer $t + 1$. That is, 

\begin{align}
&\left(\begin{array}{c}
p_{\circ_j,\circ_{j+1}}^{t+1} \\
p_{\circ_j,\bullet_{j+1}}^{t+1} \\
p_{\bullet_j,\circ_{j+1}}^{t+1} \\
p_{\bullet_j,\bullet_{j+1}}^{t+1}
\end{array}\right)= 
\left(\begin{array}{cccc}
1 & 0 & 0 & 0 \\
0 & a & a & a \\
0 & a & a & a \\
0 & 1-2 a & 1-2 a & 1-2 a
\end{array}\right)\left(\begin{array}{c}
p_{\circ_j,\circ_{j+1}}^{t} \\
p_{\circ_j,\bullet_{j+1}}^t \\
p_{\bullet_j,\circ_{j+1}}^t \\
p_{\bullet_j,\bullet_{j+1}}^t
\end{array}\right)
\end{align}
where $a = \frac{q^2 - 1}{q^4 - 1} = \frac{1}{q^2 + 1}$. 
As discussed, the error channels can be traded for a depolarizing noise with an appropriately chosen noise parameter $f$. The effect of noise can thus be accounted for by the multiplication of $f^w$ after each unitary layer, where $w$ is the weight of the Pauli operator. This factor can be included in the time evolution by multiplying  $(\circ_j\bullet_{j+1})$ and $(\bullet_j\circ_{j+1})$ by $f$ and $(\bullet_j\bullet_{j+1})$ by $f^2$ at each time step. One time step is thus characterized by the matrix 

\begin{align}
&T_j \equiv \left(\begin{array}{cccc}
1 & 0 & 0 & 0 \\
0 & f & 0 & 0 \\
0 & 0 & f & 0 \\
0 & 0 & 0 & f^2
\end{array}\right) \left(\begin{array}{cccc}
1 & 0 & 0 & 0 \\
0 & a & a & a \\
0 & a & a & a \\
0 & 1-2 a & 1-2 a & 1-2 a
\end{array}\right)  = 
\left(\begin{array}{cccc}
1 & 0 & 0 & 0 \\
0 & f a & f a & f a \\
0 & f a & f a & f a \\
0 & f^2(1-2 a) & f^2(1-2 a) & f^2 (1-2a)
\end{array}\right).
\end{align}
We denote by $p_{i,\vec{w}(t)}$ the coefficients associated with the weight vector $\vec{w}(t)$ at layer $t$. We may then find $p_{i,\vec{w}(t+1)}$ by applying the update matrix $T_{\vec{w}(t+1),\vec{w}(t)}$ on $p_{i,\vec{w}(t)}$ 
\begin{equation}
p_{i,\vec{w}(t+1)} = \sum_{\vec{w}(t)}T_{\vec{w}(t+1),\vec{w}(t)}p_{i,\vec{w}(t)}
\end{equation}
where 
\[ T_{\vec{w}(t+1),\vec{w}(t)} = \begin{cases} \prod_{j \text{ odd}}T_j & t\text{ is odd} \\ \prod_{j \text{ even}}T_j & t\text{ is even}. \end{cases} \] Finally, we note that we must multiply by an additional factor of $f^{\norm{\vec{w}_i}^2}$, with $\norm{\vec{w}_i}^2$ the weight of the initial Pauli operator $P_i$, to account for the first layer of noise.
\subsubsection{Numerical simulation}\label{sec:NumericalMPS}
We now discuss an efficient numerical algorithm to compute the eigenvalue $\beta_{i,\varepsilon}$ based on the numerical algorithm used in \cite{IPPOLITI2023_OperatorSpreading}. Naively applying the time evolution step $T_{\vec{w}(t+1), \vec{w}(t)}$ on the coefficients $p_{i,\vec{w}(t)}$ is not a viable option here, since the number of entries scales exponentially with the operator size. We expect the minimum of the shadow norm to occur at a circuit depth $t_*$ that scales maximally as $\ln(k)$ where $k$ is the number of sites that are not the identity operator in the initial Pauli operator $P_i$ \cite{IPPOLITI2023_OperatorSpreading, HUANG2022_ClassicalShadowsML, BERTONI2022_LowDepthShadows}. This implies that an MPS representation of the probability vector $\vec{p}_{i,\vec{w}(t)}$ should be an efficient way to evolve the system layer by layer. The eigenvalue $\beta_{i,\varepsilon}$  can be computed via the scalar product between the vector $\vec{\kappa}_{\vec{w}(t)}$ and the time-evolved probability vector $\vec{p}_{i,\vec{w}(t)}$ where $\vec{\kappa}_{\vec{w}(t)} = \frac{1}{(q + 1)^{\norm{\vec{w}(t)}^2}}$. The vector $\vec{\kappa}_{\vec{w}(t)}$ admits an MPS representation with the matrices $A_{\alpha,\beta}^{n_{2j - 1}} = \frac{1}{(q+1)^{n_j}}$, $B_{\beta,\alpha'}^{n_{2j}} = \frac{1}{(q+1)^{n_{2j}}}$ for any site $j$, where $n_j$ is equal to $1$ when the site $j$ is occupied, and is $0$ otherwise. The eigenvalue $\beta_{i,\varepsilon}$ is given by 
\begin{align}
\beta_{i,\varepsilon} = \sum_{\vec{w}(t),\vec{w}(0)}\vec{\kappa}_{\vec{w}(t)} \left(\sum_{\vec{w}(1),...,\vec{w}(t-1)}T_{\vec{w}(t),\vec{w}(t-1)}...T_{\vec{w}(1),\vec{w}(0)}\right)\vec{p}_{i,\vec{w}(0)}.
\end{align}
Now, since the vector $\vec{\kappa}_{\vec{w}(t)}$ is fixed, it is more efficient to compute the action of the update matrices on $\vec{\kappa}_{\vec{w}(t)}$ and then compute the scalar product with $\vec{p}_{i,\vec{w}(0)}$. Thus, we want to compute
\begin{equation}
\vec{\kappa}_{\vec{w}(0)} = \vec{\kappa}_{\vec{w}(t)}\left(\sum_{\vec{w}(1),...,\vec{w}(t)}T_{\vec{w}(t),\vec{w}(t-1)}...T_{\vec{w}(1),\vec{w}(0)}\right)
\end{equation}
from which we have that 
\begin{equation}
\beta_{i,\varepsilon} = \sum_{\vec{w}(0)}\vec{\kappa}_{\vec{w}(0)}\vec{p}_{i,\vec{w}(0)}.
\end{equation}
Now since the product $\left(\sum_{\vec{w}(1),...,\vec{w}(t)}T_{\vec{w}(t),\vec{w}(t-1)}...T_{\vec{w}(1),\vec{w}(0)}\right)$ is invariant under translation by two sites and the vector $\vec{\kappa}_{\vec{w}(t)}$ is translationally invariant, the vector $\vec{\kappa}_{\vec{w}(j)}$ for all $j$ can always be represented using a two-site unit cell with two matrices $A^{m_{2j-1}}_{\alpha,\beta}$, $B^{m_{2j}}_{\beta,\alpha'}$ where $A,B$ do not depend on the site $j$. Under the first non-trivial unitary layer, the matrices get updated as 
\begin{align}
C_{\alpha,\alpha'}^{n_{2j-1}, n_{2j}} =  \sum_{m_{2j-1},m_{2j},\beta} T_{2j-1, (n_{2j-1}, n_{2j}),(m_{2j-1}, m_{2j})} A_{\alpha,\beta}^{m_{2j-1}}B_{\beta, \alpha'}^{m_{2j}}.
\end{align}
where the pairs $(n_{2j-1},n_{2j}), (m_{2j-1},m_{2j})$ correspond to the possible weight configurations on the sites $2j-1,2j$, which are $\circ \circ$, $\circ \bullet$, $\bullet \circ$ and $\bullet \bullet$.
The resulting tensor $C_{\alpha,\alpha'}^{n_{2j-1}, n_{2j}}$ can then be written as a tensor contraction by using an SVD decomposition. We get 
\begin{equation}
C_{\alpha,\alpha'}^{n_{2j-1}, n_{2j}} \rightarrow \sum_{\beta} A'^{n_{2j-1}}_{\alpha,\beta}B'^{n_{2j}}_{\beta,\alpha'}.
\end{equation}
This provides an update rule for the matrices $A, B$. To apply the second layer of the time evolution step, we follow a completely analogous procedure, but taking into account that the circuit is shifted by one site. We get
\begin{align}
C_{\beta',\beta}^{n_{2j}, n_{2j+1}} = \sum_{m_{2j},m_{2j+1},\alpha}T_{2j,(n_{2j}, n_{2j+1}),(m_{2j}, m_{2j+1})} B_{\beta',\alpha}'^{m_{2j}}A_{\alpha, \beta}'^{m_{2j+1}}
\end{align}
\begin{equation}
C_{\beta',\beta}^{n_{2j}, n_{2j+1}} \rightarrow \sum_{\alpha}B_{\beta',\alpha}''^{n_{2j}}A_{\alpha, \beta}''^{n_{2j+1}}.
\end{equation}
To keep the computational cost low, we fix a maximum bond dimension $\chi$ for the SVD decomposition \cite{IPPOLITI2023_OperatorSpreading} of $\chi \approx 2000$ which we find is large enough to achieve numerical convergence at the cricuit depths $t$ we study. The above steps are then repeated according to the number of layers which yields an MPS representation of the final vector $\vec{\kappa}_{\vec{w}(0)}$. The final numerical task is to compute the eigenvalue $\beta_{i,\varepsilon}$. Assuming that the initial Pauli operator $P_i$ has support within a contiguous region $A$ and that $w_j$ is equal to $\bullet$ if $P_i$ is occupied at site $j$, and is $\circ$ otherwise, we find that
\begin{align}
\sum_{\vec{w}(0)}\vec{\kappa}_{\vec{w}(0)}\vec{p}_{i,\vec{w}(0)} = \lim_{r \rightarrow \infty}\operatorname{Tr}\left[(A^{\circ}B^{\circ})^{r}\prod_{(2j-1,2j) \cap A \neq 0}A^{w_{2j-1}}B^{w_{2j}}\right].
\end{align}
where the product is ordered from the smallest to the largest $j$ satisfying the condition $(2j-1,2j) \cap A \neq 0$. Now, due to the $r \rightarrow \infty$ limit, only the pair of right and left eigenvectors $\bra{E_l}, \ket{E_r}$ associated with the $+1$ eigenvalue of $A^{\circ}B^{\circ}$ survive \cite{IPPOLITI2023_OperatorSpreading}, which yields
\begin{equation}
\beta_{i,\varepsilon} = \bra{E_l}\prod_{(2j-1,2j) \cap A \neq 0 }A^{w_{2j-1}}B^{w_{2j}}\ket{E_r}
\end{equation} from which the shadow norm can be computed. 
\end{widetext}

\end{document}